\documentclass[aps,showpacs,twocolumn]{revtex4}

\usepackage{epsfig}
\usepackage{amsmath}
\usepackage{amsfonts}
\usepackage{amssymb}
\usepackage{url}
\usepackage{times}

\DeclareMathOperator{\diff}{\textrm{d}\!}

\graphicspath{{pics/}}

\begin{document}

\title{Numerical evolutions of a black hole-neutron star system 
in full
       General Relativity: \\ I. Head-on collision}

\author{Frank~L\"offler}
\affiliation{SISSA, International School for
             Advanced Studies and INFN, Via Beirut 2, 34014 Trieste, Italy}
\affiliation{Max-Planck-Institut f\"ur Gravitationsphysik,
             Albert-Einstein-Institut, 14476 Potsdam, Germany}

\author{Luciano~Rezzolla}
\affiliation{Max-Planck-Institut f\"ur Gravitationsphysik,
             Albert-Einstein-Institut, 14476 Potsdam, Germany}
\affiliation{SISSA, International School for
             Advanced Studies and INFN, Via Beirut 2, 34014 Trieste, Italy}
\affiliation{Department of Physics, Louisiana State University, Baton
             Rouge, LA 70803 USA}

\author{Marcus~Ansorg}
\affiliation{Max-Planck-Institut f\"ur Gravitationsphysik,
Albert-Einstein-Institut, 14476 Golm, Germany}

\date{June 23, 2006}

\begin{abstract}
	We present the first simulations in full General Relativity of
        the head-on collision between a neutron star and a black hole
        of comparable mass. These simulations are performed through
        the solution of the Einstein equations combined with an
        accurate solution of the relativistic hydrodynamics equations
        via high-resolution shock-capturing techniques. The initial
        data is obtained by following the York-Lichnerowicz conformal
        decomposition with the assumption of time symmetry. Unlike
        other relativistic studies of such systems, no limitation is
        set for the mass ratio between the black hole and the neutron
        star, nor on the position of the black hole, whose apparent
        horizon is entirely contained within the computational
        domain. The latter extends over $\sim 400~M$ and is covered
        with six levels of fixed mesh refinement. Concentrating on a
        prototypical binary system with mass ratio $\sim 6$, we find
        that although a tidal deformation is evident the neutron star
        is accreted promptly and entirely into the black hole. While
        the collision is completed before $\sim 300~M$, the evolution
        is carried over up to $\sim 1700~M$, thus providing time for
        the extraction of the gravitational-wave signal produced and
        allowing for a first estimate of the radiative efficiency of
        processes of this type.
\end{abstract}


\maketitle

\section{Introduction}
\label{sec:introduction}
Binary systems of compact bodies are believed to be among the major
sources of gravitational waves, which could be detected with the newly
built gravitational wave detectors. These compact objects can be black
holes or very compact stars, like neutron stars.  There is strong
evidence that objects like neutron stars or black holes do
exist~\cite{Belczynski:2001uc} and even collide
~\cite{Kalogera04,Bethe1998} in our universe.  Amongst the
astrophysical phenomena that are the best candidates for producing
detectable wave signals for the Earth-based detectors, those including
highly relativistic matter near black holes stand out.  Black
hole-neutron star binaries are believed to be as likely to be observed
as binary neutron star mergers (see, for instance,~\cite{Bethe1998}),
with expected event rates of one per year in a sphere of about
$70\mbox{Mpc}$ radius~\cite{Bethe1998} and an expected detection rate
of more than one event per year~\cite{Belczynski:2001uc} (for LIGO
II). Gravitational-wave signals from binary neutron star and mixed
binary systems in close orbit are expected to give information about
the neutron star structure and equation of state
(EOS)~\cite{Belczynski:2001uc,Bejger05}. However calculating these
gravitational waves is far too difficult to be done analytically and
numerical investigations are the only possible avenue in the final
stages of the evolution, when nonlinear effects and finite-size
contributions from the compact star need to be modeled accurately. The
importance of such an accurate modeling, however, is that the
analysis of the gravitational-wave signal could be used to obtain
important information about the equation of state governing the matter
of the compact star~\cite{Vallisneri00}.

Mixed binaries are also interesting for a number of other reasons,
\textit{e.g.} they could be a source of short gamma-ray bursts
(\textit{e.g.}~\cite{Narayan92,Fox2005}), that last typically for
about a fraction of a second~\cite{Kouveliotou1993}. Previous studies
of the the dynamics of mixed binary systems in Newtonian gravity and
carried out either analytically~\cite{Zwart1998a}, or through
numerical simulations~\cite{Lee99a,Lee99b}, suggest that a stable or
repeated, but in any case long-lasting, mass transfer from the neutron
star to the black hole is to be expected. However that might be
different in General Relativity~\cite{Miller05}, which is required in
the presence of black holes. Considerable effort has gone into
modeling systems of binaries containing either two black holes
(\textit{e.g.}
\cite{Campanelli:2006gf,Diener-etal-2006a,Baker:2006yw} and references
therein) or two neutron stars (\textit{e.g.}
\cite{Shibata:2003ga,Shibata06a} and references therein).

Research on {\it mixed} binary systems has traditionally been
performed using Newtonian gravity. In particular, together with some
analytical work on stable mass transfer from a neutron star to a black
hole~\cite{Zwart1998a}, numerical simulations have been carried out
using smooth particle hydrodynamics (SPH) techniques and with either
soft~\cite{Lee99a} or stiff~\cite{Lee99b} equations of state. All of
this bulk of work suggests that, depending on the mass ratio, at least
part of the neutron star should be tidally disrupted before the
merger. These approaches have also provided the first estimates of the
gravitational-wave emission during the inspiral and merger by making
use of the Newtonian quadrupole approximation~\cite{Lee99a,Lee99b} and
indicating that the process has in general an efficiency of $\approx0.7\%$.
Shortly after, simulations by Janka et al.~\cite{Janka99a} included a
realistic EOS and showed good agreement with this estimate.

More recently these studies have been improved with either the use of
relativistic corrections to the Newtonian gravity in terms of
pseudo-Newtonian potentials~\cite{Rosswog05} or in General Relativity
but within a ``quasi-stationary''
approximation~\cite{Baumgarte:2004}. The latter assumes that the
evolution of the binary system can be modeled through a sequence of
quasi-stationary spacetimes, solutions of the Einstein equations.
This is clearly an approximation but a
very good one especially in those stages of the evolution of the
system when the inspiral timescale is much longer than the orbital or
the one associated to a dynamical instability. 

However, as the separation between the two compact objects decreases,
this approximation becomes less accurate and eventually
breaks~\cite{Alcubierre2003:pre-ISCO-coalescence-times,Miller04},
forcing the use of fully dynamical simulations. A first step in this
direction within a relativistic context has been made recently within
the characteristic formulation of Einstein
equations~\cite{Bishop05}. In this approach, it was shown that the
initial spurious gravitational waves signal resulting from the
approximations in the calculation of the initial data are rapidly
radiated away and the binary system relaxes to a quasi-equilibrium
state with an approximate orbital motion of the star which has been
followed only for a small fraction of an orbit.

A second step towards a more accurate description of the dynamics of a
mixed binary system has been made by Faber et al.~\cite{Faber05}. Taking
as initial data the one computed in ref.~\cite{Baumgarte:2004}, Faber et
al. have studied the final dynamics of the binary system using SPH
techniques for the hydrodynamics and a conformally flat approximation to
the Einstein equations. In these calculations the black hole is only
included as a background metric in addition to the one of the neutron
star and it is not included in the computational domain, thus limiting
the investigation to rather large black hole-to-neutron star mass
ratios. Overall, the results found indicate that rather generically, an
accretion disc can be formed from the tidal disruption of the compact
star and that this could, although short-lived, provide energy to power a
gamma-ray burst. As an extension to the results presented in
ref.~\cite{Faber05}, Taniguchi et al. have considered the evolution of
the system in the case in which the binary is
irrotational~\cite{Taniguchi05}, and found that the effect of the spin of
the neutron star has only a minor effect on the location of the tidal
break-up which plays an important role in the form of the gravitational
waves produced.  Sopuerta et al. took another
approach~\cite{Sopuerta:2006bw} to the full problem. Here, the spacetime
is treated in full General Relativity, however, the hydrodynamics of the
neutron star is frozen. This is a good approximation if the dynamical
timescales related with the deformation of the neutron star are much
bigger than the orbital timescales and also aims at large black
hole-to-neutron star mass ratios.

While important first steps in the study of the dynamics of mixed
binary systems in General Relativity, the analysis carried out in
refs.~\cite{Bishop05,Faber05,Taniguchi05,Sopuerta:2006bw}
was not able to investigate
binaries with comparable mass. This regime is far more demanding from
a computational point of view as it requires the solution of the full
Einstein equations, of accurate hydrodynamical techniques and, most
importantly, the direct inclusion of the black-hole's apparent horizon
within the computational domain. In this
context, the expectation is that the merger is prompt in most
cases~\cite{Miller05} with the formation of an accretion disc being
very unlikely. This is expected to be the result of the intense
angular-momentum loss due to gravitational radiation, which causes a
direct merger rather than a slower and extended mass transfer.

Clearly, fully numerical calculations are needed to confirm this
expectation and in this spirit we here consider the first simulations
in full General Relativity of the head-on collision between a neutron
star and a black hole of comparable mass.  We solve the Einstein
equations combined with an accurate solution of the relativistic
hydrodynamics equations via high-resolution shock-capturing (HRSC)
techniques. The initial data is obtained by following the
York-Lichnerowicz conformal decomposition with the assumption of time
symmetry, using a spectral solver on one, compactified domain. Unlike
the previously mentioned relativistic studies of such systems, no
limitation is set for the mass ratio between the black hole and the
neutron star, nor on the position of the black hole, whose apparent
horizon is entirely contained within the computational domain. The
latter extends over $\sim 400 M$ and is covered with six levels of
fixed mesh refinements.  We concentrate on an example of a binary
system with mass ratio $\sim 6$, and find that although a tidal
deformation is evident the neutron star is accreted promptly and
entirely into the black hole. The collision is completed before a time
of $\sim 300~M$, but we are able to obtain an evolution up to $\sim
1700~M$, which provides enough time for the extraction of the
gravitational-wave signal produced and allows for a first estimate of
the radiative efficiency of processes of this type.

This paper is organized as follows: Section~\ref{equations} describes the
formulation used for the Einstein and hydrodynamics equations. The
construction and testing of initial data for the mixed binary system is
presented in Section~\ref{sec:initial_data}. Section~\ref{sec:evolution}
is devoted to discussions about improved gauges and the use of
dissipation together with excision, as well as the study of the dynamics
of the merger, which depends on these techniques. Gravitational waves are
one of the main motivations for this study and in Section~\ref{gwe} we
show the first waveform obtained in a simulation in full General
Relativity of a head-on collision of a mixed binary of comparable mass
and give an order-of-magnitude estimate of the energy emitted in
gravitational waves. Finally, we conclude and outline our future plans in
Section~\ref{conclusion}.

Throughout this paper Greek indices run from $0$ to $3$ indicating the
four dimensions of spacetime and $0$ referring to the direction of
time. Unless otherwise stated, Latin indices run from $1$ to $3$.  In all
formul\ae~indices occurring twice are to be summed over the possible
range for that index.  Unless stated otherwise, we adopt the convention of
units in which $c=G=M_{\odot}=1$. Where we measure times and lengths in units
of $M$, we define $M\equiv M_\odot$.


\section{Basic equations and their implementation}
\label{equations}

        The {\tt Whisky} code solves the general relativistic
hydrodynamics equations on a 3D numerical grid with Cartesian
coordinates~\cite{Baiotti03a,Baiotti04}. The code has been constructed
within the framework of the {\tt Cactus} Computational Toolkit, developed at
the Albert Einstein Institute
(Golm) and at the Louisiana State University (Baton Rouge). While the
{\tt Cactus} code provides at each timestep a solution of the Einstein
equations~\cite{Alcubierre99d}
\begin{equation}
\label{efes}
G_{\mu \nu}=8\pi T_{\mu \nu}\ , 
\end{equation}
where $G_{\mu \nu}$ is the Einstein tensor and $T_{\mu \nu}$ is the
stress-energy tensor, the {\tt Whisky} code provides the time evolution
of the hydrodynamics equations, expressed through the conservation
equations for the stress-energy tensor $T^{\mu\nu}$ and for the matter
current density $J^\mu$
\begin{equation}
\label{hydro eqs}
\nabla_\mu T^{\mu\nu} = 0\;,\;\;\;\;\;\;
\nabla_\mu J^\mu = 0\ .
\end{equation}
Hereafter we will consider the dynamics of perfect fluids described by a
stress-energy tensor 
\begin{equation}
T^{\mu\nu} = \rho h u^{\mu} u^{\nu} + p g^{\mu\nu}\ ,
\end{equation}
where $h\equiv 1 +\epsilon + {p}/{\rho}$ is the specific enthalpy, while
$p$, $\rho$ and $\epsilon$ indicate the pressure, the rest-mass density
and the internal energy density, respectively. Furthermore, we will
assume the fluid to obey an ``ideal fluid'' ($\Gamma$-law) equation of
state
\begin{equation}
\label{eq:EOSideal}
p = (\Gamma-1) \rho\, \epsilon \ ,
\end{equation}
which, for $\Gamma=2$, provides a reasonably good approximation for a
stiff equation of state.

        In what follows we briefly discuss how both the right and the
left-hand side of equations (\ref{efes}) are computed within the coupled
{\tt Cactus/Whisky} codes.

\subsection{Evolution of the field equations}   
\label{feqs}

        We here give only a brief overview of the system of equations for
the evolution of the field equations, but refer the reader
to~\cite{Alcubierre99d} for more details. Many different formulations of
the equations have been proposed throughout the years, starting with the
ADM formulation in 1962~\cite{Arnowitt62}. As mentioned in the
Introduction, we use the conformal traceless
formulation~\cite{Nakamura87}, which is based on the ADM construction and
has been further developed in~\cite{Shibata95}.

Details of our particular implementation are extensively described
in~\cite{Alcubierre99d,Alcubierre02a} and will not be repeated here. We
only recall that this formulation makes use of a conformal decomposition
of the three-metric, \hbox{$\tilde \gamma_{ij} = e^{- 4 \phi}
\gamma_{ij}$}, and the trace-free part of the extrinsic curvature,
\hbox{$A_{ij} = K_{ij} - \gamma_{ij} K/3$}, with the conformal factor
$\phi$ chosen to satisfy $e^{4 \phi} = \gamma^{1/3}$, where $\gamma$ is
the determinant of the spatial three-metric $\gamma_{ij}$. In this
formulation, in addition to the evolution equations for the conformal
three-metric $\tilde \gamma_{ij}$ and for the conformal traceless
extrinsic curvature $\tilde A_{ij}$, there are evolution equations for
the conformal factor $\phi$, for the trace of the extrinsic curvature $K$
and for the ``conformal connection functions'' $\tilde \Gamma^i \equiv
\tilde \gamma^{ij}{}_{,j}$. We note that although the final mixed,
first-order in time and second-order in space, evolution system for
$\phi, K, \tilde \gamma_{ij}, {\tilde A_{ij}}, {\tilde \Gamma^i}$ is not
in any immediate sense hyperbolic, there is evidence showing that the
formulation is at least equivalent to a hyperbolic
system (\textit{e.g.}~\cite{Bona:2003qn,Nagy:2004td}).

\subsubsection{Standard Gauge choices}

        The code is designed to handle arbitrary shift and lapse
conditions, which can be chosen as appropriate for a given spacetime
simulation.  More information about the possible families of spacetime
slicings which have been tested and used with the present code can be
found in~\cite{Alcubierre99d,Alcubierre01a}. Here, we limit ourselves to
recalling details about the specific foliations used in the present
evolutions. In particular, we have used hyperbolic $K$-driver slicing
conditions of the form
\begin{equation}
(\partial_t - \beta^i\partial_i) \alpha = - f(\alpha) \;
\alpha^2 (K-K_0),
\label{eq:BMslicing}
\end{equation}
with $f(\alpha)>0$ and $K_0 \equiv K(t=0)$. This is a generalization of
many well known slicing conditions.  For example, setting $f=1$ we
recover the ``harmonic'' slicing condition, while, by setting
\mbox{$f=q/\alpha$}, with $q$ an integer, we recover the generalized
``$1+$log'' slicing condition~\cite{Bona94b}.  In particular, all of the
simulations discussed in this paper are done using condition
(\ref{eq:BMslicing}) with $f=2/\alpha$. This choice has been made mostly
because of its computational efficiency, but we are aware that ``gauge
pathologies'' could develop with the ``$1+$log''
slicings (see \textit{e.g.}~\cite{Alcubierre97b}).

        As for the spatial gauge, we use one of the ``Gamma-driver''
shift conditions proposed in~\cite{Alcubierre01a}, that essentially act so
as to drive the
$\tilde{\Gamma}^{i}$ to be constant. More specifically, all of the
results reported here have been obtained using a (modified) hyperbolic
Gamma-driver condition,
\begin{equation}
\partial^2_t \beta^i = F \, \partial_t \tilde\Gamma^i - \eta \,
\partial_t \beta^i,
\label{eq:hyperbolicGammadriver}
\end{equation}
where $F$ and $\eta$ are, in general, positive functions of space and
time (We typically choose $F=3/4$ and $\eta=3$ and do not vary them in
time.). For the hyperbolic Gamma-driver conditions it is crucial to add a
dissipation term with coefficient $\eta$ to avoid strong oscillations in
the shift. Experience has shown that by tuning the value of this
dissipation coefficient it is possible to almost freeze the evolution of
the system at late times. As mentioned in~\cite{Baiotti04}, we here
recall that the ``Gamma-driver'' shift conditions are similar to the
``Gamma-freezing'' condition $\partial_t \tilde\Gamma^k=0$, which, in
turn, is closely related to the well-known minimal distortion shift
condition~\cite{Smarr78b}. The differences between these two conditions
involve the Christoffel symbols and are basically due to the fact that
the minimal distortion condition is covariant, while the Gamma-freezing
condition is not.

\subsection{Evolution of the hydrodynamics equations}    

        An important feature of the {\tt Whisky} code is the
implementation of a \textit{conservative formulation} of the
hydrodynamics equations~\cite{Marti91,Banyuls97,Ibanez01}, in which the
set of equations (\ref{hydro eqs}) is written in a hyperbolic,
first-order and flux-conservative form of the type
\begin{equation}
\label{eq:consform1}
\partial_t {\mathbf q} + 
        \partial_i {\mathbf f}^{(i)} ({\mathbf q}) = 
        {\mathbf s} ({\mathbf q})\ ,
\end{equation}
where ${\mathbf f}^{(i)} ({\mathbf q})$ and ${\mathbf s}({\mathbf q})$
are the flux-vectors and source terms, respectively~\cite{Font03}.  Note
that the right-hand side of eq.~\eqref{eq:consform1}, {\it i.e.,} the
source vector ${\mathbf s}({\mathbf q})$, depends only on the metric, and
its first derivatives, and on the stress-energy tensor. Furthermore,
while the system (\ref{eq:consform1}) is not strictly hyperbolic, strong
hyperbolicity is recovered in a flat spacetime, where ${\mathbf s}
({\mathbf q})=0$.

\begin{figure*}
 \centering
 \includegraphics[width=8.cm]{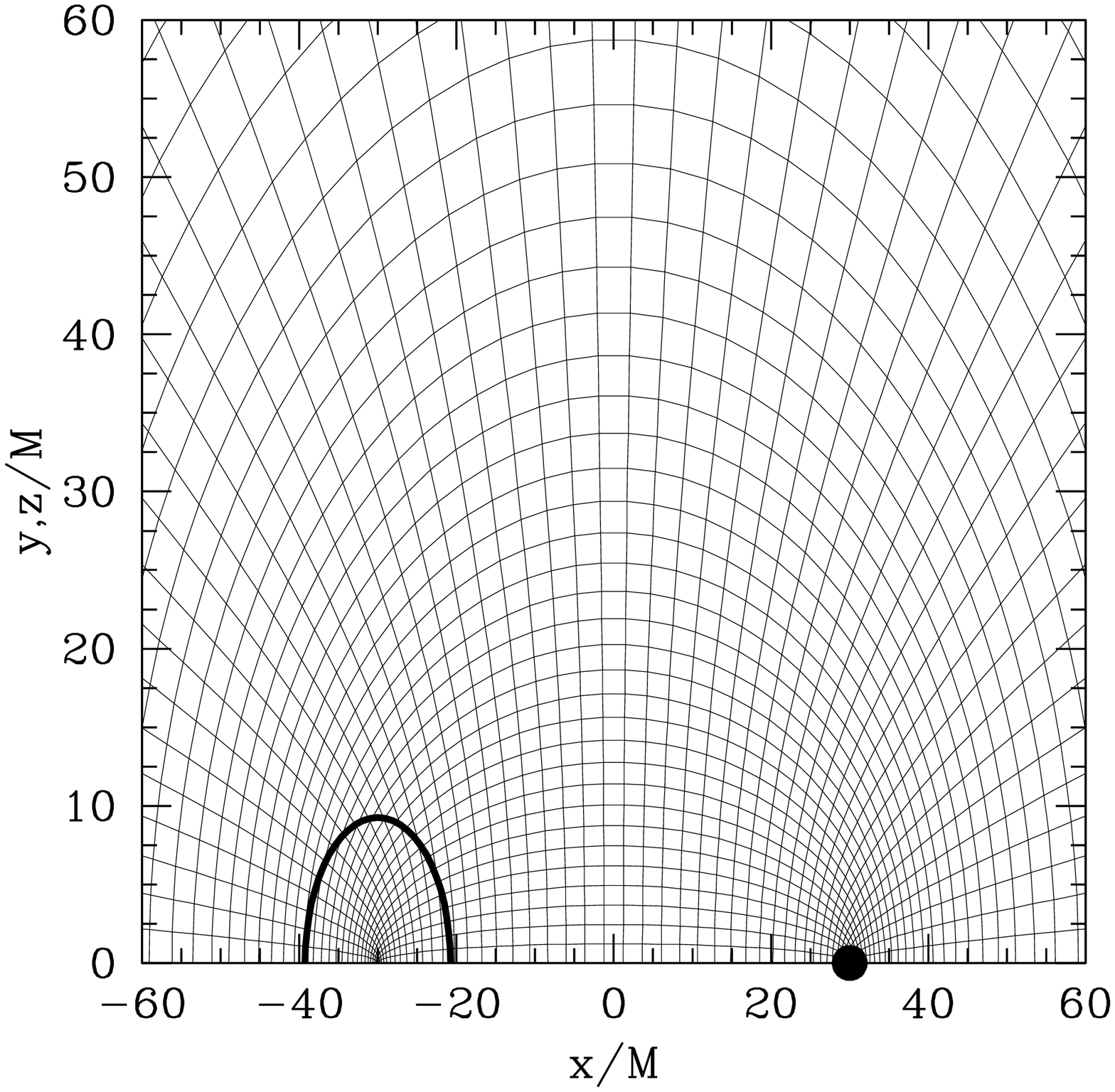}
 \hskip 1.0cm
 \includegraphics[width=8.cm]{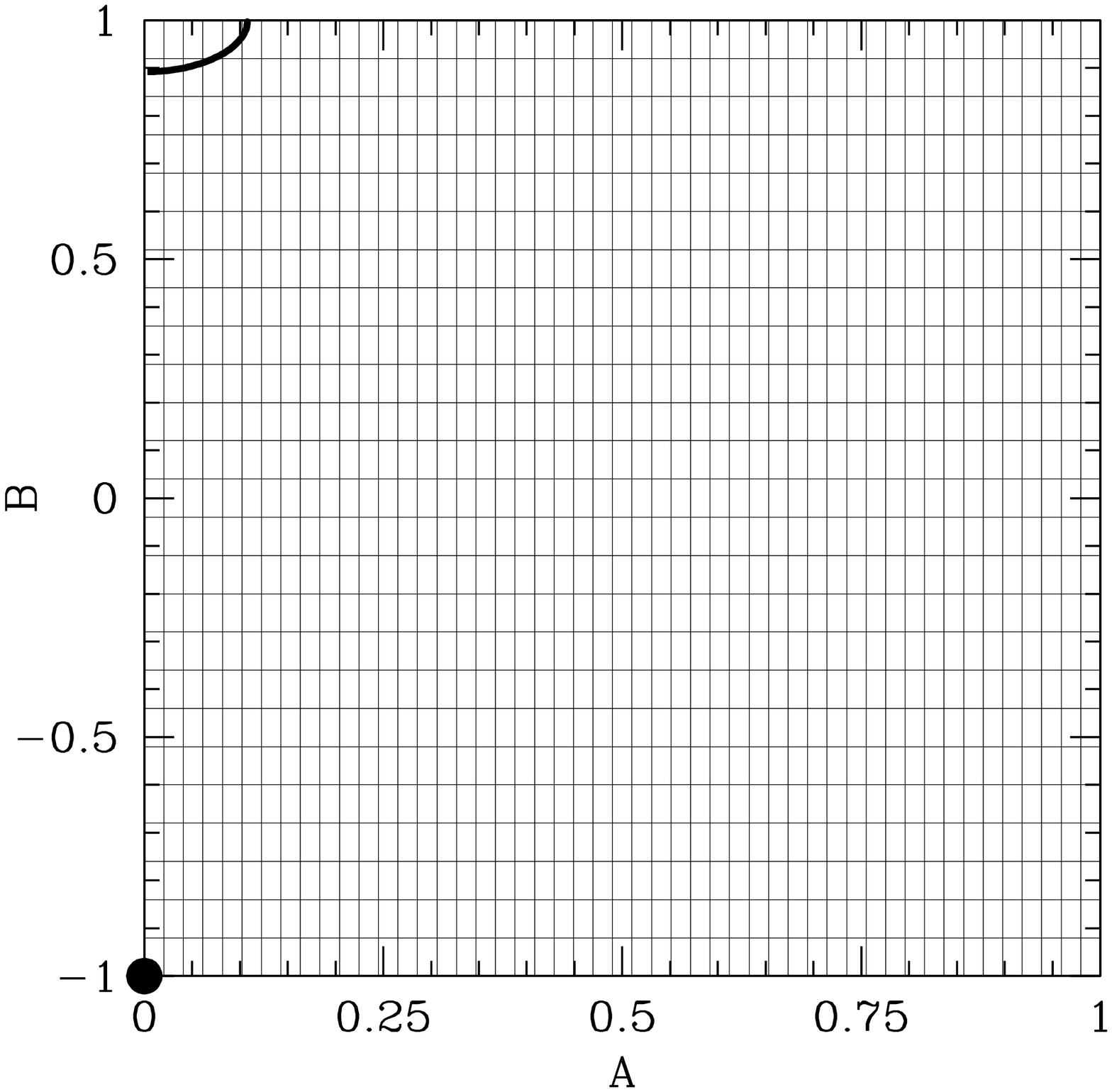}\\
 \caption{{\it Left panel:} Coordinate lines of $A$ and $B$ on the
          Cartesian grid. The arrow denotes the location of the
          singularity ($b=30$) and the sector is a sketch of the neutron
          star. {\it Right panel:} coordinate lines of $A$ and $B$ on
          their grid. The arrow denotes the location of the singularity
          and the sector is a sketch of the neutron star.}
 \label{fig:coord_1}
\end{figure*}

        Additional details of the formulation we use for the
hydrodynamics equations can be found in~\cite{Baiotti04}. We stress that
an important feature of this formulation is that it has allowed to extend
to a general relativistic context the powerful numerical methods
developed in classical hydrodynamics, in particular high-resolution
shock-capturing schemes based on
linearized Riemann solvers (see~\cite{Font03} for a extensive discussion
of this). Such schemes are essential for a correct representation of
shocks, whose presence is expected in several astrophysical scenarios.

\section{Initial Data for Mixed Binaries}
\label{sec:initial_data}

\subsection{A Spectral Approach for  Mixed Binaries}
\label{sec:spectral_id}

The initial data for a binary system composed of a black hole and of a
relativistic star of comparable masses have been computed with a modified
version of the spectral solver used in ref.~\cite{Ansorg:2004ds} for the
solution of the initial data problem in a binary system of two black holes
modeled as punctures~\cite{Brandt00}. While we refer the interested
reader to~\cite{Ansorg:2004ds} for the details of the solver, we here
concentrate on the modifications needed in order to include matter
sources in the initial slice of the spacetime. 

In a vacuum spacetime describing two black holes with zero momenta, the
only non-trivial equation is the Hamiltonian constraint, which can be
written as
\begin{equation}
 \label{eq:ham_vacuum}
 \tilde{\nabla}^2\psi-\frac{1}{8}\psi\tilde{R}-\frac{1}{8}\psi^5K^2+
 \frac{1}{8}\psi^5K_{ij}K^{ij}=0 \ .
\end{equation}
where $\tilde{\nabla}^2$ and ${\tilde R}$ are the Laplace operator and
the Ricci scalar connected to the conformal metric, respectively.  If
a non-vacuum configuration is considered, as it is necessary for mixed
binary systems, eq. \eqref{eq:ham_vacuum} needs to be extended to
\begin{equation}
 \label{eq:ham_matter_full}
 \tilde{\nabla}^2\psi-\frac{1}{8}\psi\tilde{R}-\frac{1}{8}\psi^5K^2+
 \frac{1}{8}\psi^5K_{ij}K^{ij}=-2\pi\psi^5 e_{_{\rm ADM}}  \ ,
\end{equation}
where
\begin{equation}
\label{eq:e_ADM}
e_{_{\rm ADM}} = n^\mu n^\nu T_{\mu\nu}
                         = W^2\left(\rho(1+\epsilon)+p\right)-p \ ,
\end{equation}
is the energy density measured by an observer moving orthogonally to the
spacelike hypersurface with normal unit vector ${\boldsymbol n}$ and $W$ the
Lorentz factor.

It can be shown that it is not possible to solve
eq.~\eqref{eq:ham_matter_full} directly using $e_{_{\rm ADM}}$ since
this does not produce a solution in general~\cite{York79}. Rather, it
is necessary to adopt a new quantity ${\bar e}$ related to $e_{_{\rm
    ADM}}$ as
\begin{equation}
 \label{eq:e_bar}
 {\bar e} = \psi^n e_{_{\rm ADM}} \ ,
\end{equation}
where the exponent $n$ is arbitrary but for the condition $n>5$ (see
ref.~\cite{York79} for the different possible choices). We have used
$n=8$ so that eq.~\eqref{eq:ham_matter_full} becomes
\begin{equation}
 \label{eq:ham_matter2}
 \tilde{\nabla}^2\psi-\frac{1}{8}\psi\tilde{R}-\frac{1}{8}\psi^5K^2+
 \frac{1}{8}\psi^5K_{ij}K^{ij}=-2\pi\psi^{-3}{\bar e}  \ .
\end{equation}
The quantity $\bar e$ is first computed using eq.~\eqref{eq:e_bar}
from the conformal factor $\psi$ and from the initial guess for the
fluid quantities ($p, \rho$ and $\epsilon$), and then held fixed while
solving eq.~\eqref{eq:ham_matter2}. From $\bar e$ and the new solution
for $\psi$, the new $e_{_{\rm ADM}}$ can be obtained again through
eq.~\eqref{eq:e_bar} and, as a result, also the new values for the
fluid quantities using eq.~\eqref{eq:e_ADM} and the polytropic EOS
with $\Gamma=2$. As initial guess we used the solution for a spherically
symmetric star in equilibrium.

An accurate solution to eq.~\eqref{eq:ham_matter2} can be found using
spectral methods and a set of coordinates $(A, B, \phi)$ which are
suitably adapted to the geometry of the problem and which are related
to the Cartesian ones through the transformation
\begin{align}
 \label{eq:coordinate_transformation}
 x &= \frac{2B(A^2+1)}{(A^2-1)(1+B^2)}b,\\
 y &= -\frac{2A(1-B^2)}{(A^2-1)(1+B^2)}b\cos\phi,\\
 z &= \frac{2A(1-B^2)}{(1-A^2)(1+B^2)}b\sin\phi,
\end{align}
where $b$ is half of the separation between the two singularities, $\phi$ is the
standard azimuthal coordinate in a spherical polar coordinate system, and
the coordinates $A$ and $B$ are restricted to be
\begin{equation}
 A\in[0,1],\quad B\in[-1,1],\quad \phi\in[0,2\pi) \ .
\end{equation}
Fig.~\ref{fig:coord_1} provides a schematic picture of the coordinate
setup and the location of the star and of the black hole in both spaces
of coordinates.

Following the puncture approach, we decompose $\psi$ into a part which is
a known solution to eq.~\eqref{eq:ham_matter_full} and contains the
contribution of the single black hole, and an unknown part $u$
\begin{equation}
 \label{eq:phi_u}
 \psi=1+\frac{m}{2r}+u \ .
\end{equation}
As a result, eq.~\eqref{eq:ham_matter_full} reduces to an equation of the
type
\begin{equation}
 \label{eq:basic_eq_for_u}
 f(u) = \tilde{\nabla}^2 u = 0 \ .
\end{equation}
Clearly, eq.~\eqref{eq:basic_eq_for_u} does not have any singular
behavior and can be computed very accurately as discussed in
ref.~\cite{Ansorg:2004ds}.  Note that in the case of a vacuum
spacetime the momentum constraint equation can be solved analytically
and thus the simplifying assumption of zero-momenta can be dropped. When
matter sources are present, on the other hand, this is no longer
possible. Because of the added complications coming from this,
hereafter we will restrict ourselves to considering the simplest case
in which both the black hole and the compact star to have zero
momenta, {\it i.e.,} $j^i=0$. As a result eq.~\eqref{eq:ham_matter2}
reads
\begin{equation}
 \label{eq:ham_matter}
 \tilde{\nabla}^2\psi = -2\pi\psi^{-3}{\bar e} \ .
\end{equation}

\begin{figure}
 \centering \includegraphics[width=8.5cm]{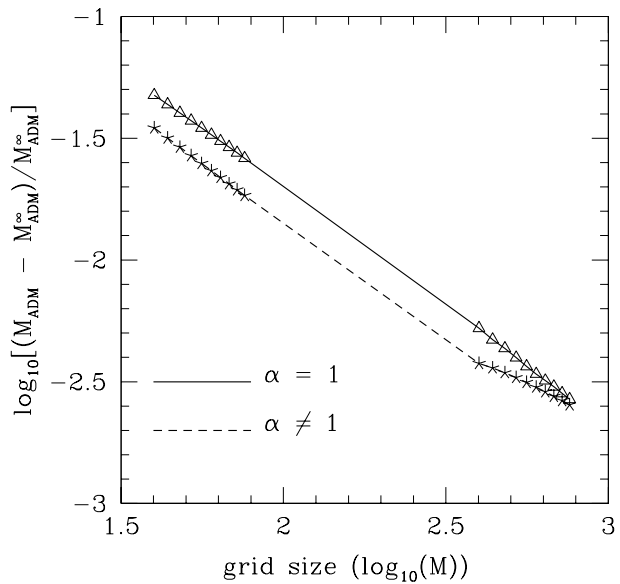}\\
 \caption{Fractional error in the evaluation of the ADM mass resulting
          from a finite size computational domain. The two curves show
          the error in the different cases in which the lapse function in
          eq.~\eqref{eq:ADM_mass} is set to be one or not.}
 \label{fig:madm_vs_size}
\end{figure}

	The use of the coordinate
system~\eqref{eq:coordinate_transformation} has also the important
advantage that it compactifies the spatial domain and thus allows for an
accurate measurement of the ADM mass on the spacelike slice
\begin{align}
 \label{eq:ADM_mass_volume}
 M_{_{\rm ADM}}&=\frac{1}{16\pi}\int
 \left(\alpha\sqrt{\gamma}\gamma^{ij}\gamma^{kl}
 (\gamma_{ik,j}-\gamma_{ij,k})\right)_{,l}\diff^{\,3}\!\!x\\
 \label{eq:ADM_mass}
 &=\frac{1}{16\pi}\oint_\infty
 \alpha \sqrt{\gamma}\gamma^{ij}\gamma^{kl}
 (\gamma_{ik,j}-\gamma_{ij,k})\diff S_l.
\end{align}
where the second expression converts the volume integral to a surface
one. This latter expression is also evaluated on the grid used for the time
evolution over the largest cuboidal surface that can be fitted on the
finite-size computational domain covered by the Cartesian coordinates.
Clearly, this
represents an approximation to the actual ADM mass and the overall error
can be estimated by comparing the values of $M_{_{\rm ADM}}$ as computed
initially on the compactified, spectral grid with the one estimated on
the Cartesian grid. This fractional error is as shown as a function of
the linear grid size in Fig.~\ref{fig:madm_vs_size}, where the two curves
show the error whether or not the lapse function in
eq.~\eqref{eq:ADM_mass} is set to be one as in an asymptotically flat
spacetime. For each choice of grid size, this error will determine the
size of an ``error-bar'' indicating the smallest error for the
measurement of the ADM mass for the typical grid resolutions used in our
simulations; ({\it cf.} the error-bar in Fig.~\ref{fig:ah_mass}).

Once the solution to eq.~\eqref{eq:ham_matter2} is found with the
desired accuracy on the spectral grid, it is then evaluated on a
Cartesian grid and used as initial data for the subsequent
evolution. The evaluation on the Cartesian grid can be done by either
using the same set of spectral coefficients used for the spectral
solution (see discussion in ref.~\cite{Ansorg:2004ds}) or by simply
performing a polynomial interpolation of the desired variables from
the spectral grid onto the Cartesian one. The first approach has the
advantage of being much more accurate but also extremely costly from a
computational standpoint; furthermore, as the accuracy during the
evolution is dominated by the truncation error and not by the error
coming from the initial data, we routinely import the spectral solution
onto the Cartesian grid by means of a simple and comparatively
inexpensive third-order polynomial interpolation

It is worth noting that even in the case of compact objects of
comparable mass, a much higher resolution is needed in the vicinity of
the black hole. This is due partly to the fact that the spacetime
curvature varies there more rapidly and partly to the need of having a
sufficient resolution for the excision technique to operate
satisfactorily.
\begin{figure*}
 \vskip -7.cm
 \centering \includegraphics[width=16cm]{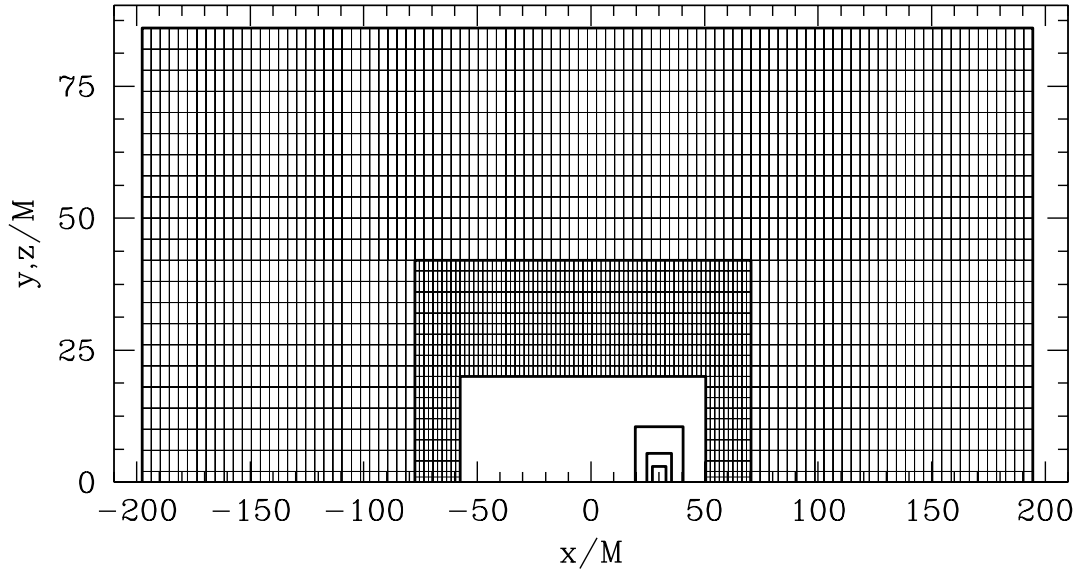}\\
 \vskip -1.5cm
 \caption{Computational setup of the mixed binary problem discussed here.
          Shown are the borders of the six mesh refinement levels (The
          actual resolution is higher than sketched here). Due to the use
          of reflection symmetry at the $x$-$y$ and the $x$-$z$ plane, we
          only have to use grid points at positive $y$ and positive $z$
          coordinates.}
 \label{fig:comp_setup}
\end{figure*}
Because the use of a high-resolution, uniform grid would be
computationally too expensive, we use a Berger-Oliger type, box-in-box,
fixed mesh-refinement~\cite{Berger-1982} as provided by the {\tt Carpet}
code~\cite{Schnetter-etal-03b}. The resulting setup of the computational
domain is sketched in Fig.~\ref{fig:comp_setup}. Note that we use a total
of six levels of mesh refinement, three of which refine the region around the
black hole. Note also that the use of
reflection symmetry across the $x$-$y$ and the $x$-$z$ planes, gridpoints
are needed only at positive $y$ and positive $z$ coordinates.

Hereafter we will concentrate on two different grid resolutions which
will refer to as ``low'' and ``high-resolution'' respectively. In the
first one we set the coarsest and finest grid spacings to be $\Delta_{\rm
coarse} =4.0~M$ and $\Delta_{\rm fine} =0.125~M$, respectively, where $M$
is the total mass of the system ($\Delta_{\rm coarse} \approx0.69~M_{\rm
ADM}$ and $\Delta_{\rm fine} \approx 0.02~M_{\rm ADM}$.). For the
high-resolution grid, on the other hand, we choose $\Delta_{\rm coarse}
=2.0~M$ and $\Delta_{\rm fine} =0.0625~M$.

\subsection{Initial-Data Testing}

The use of a spectral method for the calculation of the initial data
has the considerable advantage of providing an exponential convergence
rate when the solution is of class ${\cal C}^{\infty}$. This important
property is however lost when the solution comprises matter source
terms like in our case since these are of class ${\cal C}^0$, so that
$\psi$ is of class ${\cal C}^2$. In this case the convergence rate is
third-order in the number of coefficients $N_{\rm S}$ used in the
spectral expansion. We have tested and verified this scaling behavior
by setting the energy density on the right-hand-side of
eq.~\eqref{eq:ham_matter} to be the energy density of a spherical star
in equilibrium. Because the latter can be calculated to great accuracy
as a solution of an ordinary differential equation, all of the
truncation error is the one related to the spectral solver and we have
found it to be the expected third-order.

\begin{figure}
 \centering
 \includegraphics[width=8.5cm]{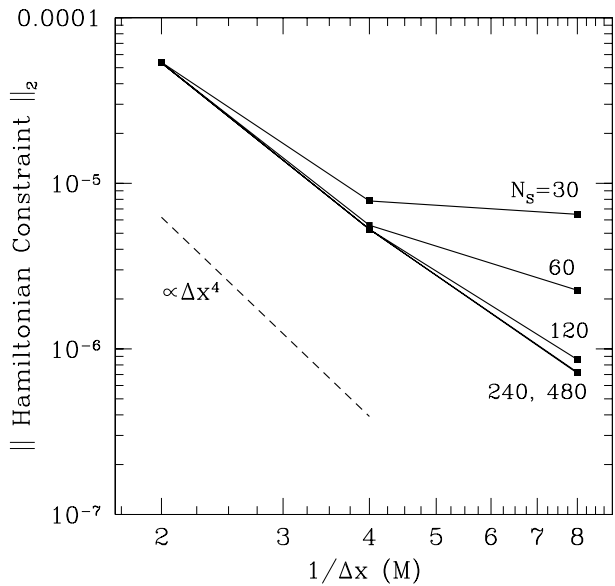}\\
 \caption{2-Norm of the Hamiltonian constraint along the $x$-axis for
   different resolutions of the spectral and the Cartesian grid.  The
   dashed line corresponds to a fourth-order convergence rate and is
   offered as a comparison.}
 \label{fig:tp_bhns_conv_ham_log}
\end{figure}

This is summarized in Fig.~\ref{fig:tp_bhns_conv_ham_log} which shows
the 2-norm of the Hamiltonian constraint as a function of the
grid-spacing $\Delta x$ and of the number of for the initial data. In
particular, different lines refer to solutions computed with the
spectral method and with a number of spectral coefficients ranging
from $N_{\rm S} = 30$ to $N_{\rm S} = 480$. The initial data is then
imported onto a Cartesian grid having a spacing ranging from $\Delta =
0.5~M$ to $\Delta = 0.125~M$ and the Hamiltonian constraint (which we
recall involves second-order spatial derivatives) is then computed
using a fourth-order accurate finite-difference stencil. Clearly, for
small Cartesian resolution ({\it i.e.,} $\Delta = 0.5~M$) the
finite-differencing error is the largest for any number of the
spectral coefficients. On the other hand, as the Cartesian
discretization decreases, the truncation error resulting from the
spectral solution becomes comparable with the finite-difference
one. As a result, more and more spectral coefficients are needed to
reach the desired fourth-order convergence and whose representative
slope is indicated with a dashed line.

Figure~\ref{fig:tp_bhns_conv_ham_log} also shows that only when a
sufficient number of spectral coefficient is used ({\it i.e.,} $N_{\rm
  S} \gtrsim 240$) is the total truncation error dominated by the
finite-differencing and that the latter is not exactly of fourth-order
but somewhat smaller ({\it i.e.,} $\approx 3.3$). The reason for this
is that the initial-data solution shows very small oscillations on the
spectral grid because of the non-differentiability of the fluid
variables at the surface of the star. Although spurious, these
oscillations in the solutions are well-resolved on the spectral grid
but cannot be satisfactorily reproduced onto the Cartesian grid,
despite the resolution being largely increased in the vicinity of the
two compact objects. As a result, the fourth-order convergence rate is
lost in these regions and it is consequently spoiled overall, bringing
it to the measured value of 3.3. 

To test the accuracy of the initial-data solver we have therefore
considered binary configurations not suffering from this problem ({\it
  e.g.}  with vacuum solutions of punctures) and recovered in those
cases the expected fourth-order convergence. We also note that while a
degraded convergence rate in the initial data is annoying, it is
rather inevitable, at least with the present choice of the domain
decomposition for the spectral grid and with the presence of matter
terms which are not of class ${\cal C}^{\infty}$. As a result, and at
least in principle, a solution to this problem could be obtained by
using multiple domains one of which would cover the matter source
entirely and whose edges would coincide with the stellar surface. In
practice, however, any additional effort in reducing the error in the
initial data would probably not translate into a direct advantage for
the evolution, since the truncation error coming from the
finite-difference operators always represents the largest contribution
to the error. In view of this, we have not pursued this effort
further.

\section{Results}
\label{sec:evolution}

\subsection{Improved Gauges and Excision}

Two important changes with respect to the evolutions discussed in
refs.~\cite{Hawke04,Baiotti04} have been introduced in this work and have
turned out to be essential in achieving the long-term stability needed to
perform the head-on collision. 

The first of these changes is the introduction of an artificial
dissipation of the Kreiss-Oliger type~\cite{Kreiss73} on the
right-hand-sides of the evolution equations for the spacetime
variables and the gauge quantities. More specifically, for any
time-evolved quantity $q$, the right-hand-side is modified with the
introduction of a term of the type
\begin{equation}
\label{dissipative_term}
{\cal L}_{\mbox{\tiny diss}}(q) = -\varepsilon\; \partial^4_x q \ ,
\end{equation}
where $\varepsilon=\varepsilon(x,\Delta x)$ is the dissipation
coefficient, it is not necessarily constant in space and it decreases
more rapidly than the truncation error associated to ${\cal L}$, {\it
  i.e.} $\varepsilon \propto \Delta x^3$.

Using a second-order finite-difference representation of the fourth-order
partial derivative, eq.~\eqref{dissipative_term} takes the form
\begin{eqnarray}
&& {\rm L}_{\mbox{\tiny diss}} (q_i)= -\frac{\varepsilon}{16}
                    \frac{\left(q_{i-2}-4q_{i-1}+6q_i-4q_{i+1}+q_{i+2}
                    \right)}{\Delta x^4} + \nonumber \\ 
&&\hskip 5.5cm		    {\cal O}(\Delta x^2)\ .
\end{eqnarray}

While the additional dissipative term is applied throughout the grid, its
magnitude will be different according to whether a gridpoint is inside or
outside the apparent horizon which, within the {\tt Cactus}
code, is obtained using the fast finder of
Thornburg~\cite{Thornburg95,Thornburg2003:AH-finding}.
More specifically, for all gridpoints
outside the apparent horizon we set $\varepsilon (x) =
\varepsilon^{\mbox{\tiny out}} \ll 1$ and a small amount of dissipation
will here serve to damp the small oscillations produced at the
mesh-refinement boundaries and which can potentially lead to instability
(see~\cite{Schnetter-etal-03b} for a complete discussion). For all
gridpoints inside the apparent horizon, on the other hand, we set
$\varepsilon (x) = \varepsilon^{\mbox{\tiny in}} \gg
\varepsilon^{\mbox{\tiny out}}$ and the transition from the two values
can be specified through a parametrized slope. The latter does not seem
to influence significantly the quality of the evolution and most of the
time a steep, linear slope is sufficient. 

The optimal values of $\varepsilon^{\mbox{\tiny out}}$ and
$\varepsilon^{\mbox{\tiny in}}$ depend on the specific problem at
hand. Clearly, excessively small values will not produce any benefit
while values that are too large will lead to numerical instabilities. For
the results presented here we have found that a long-term stability was
obtained for $\varepsilon^{\mbox{\tiny out}}=0.01$ and
$\varepsilon^{\mbox{\tiny in}}=0.1$

The second important change introduced is related to improved gauge
conditions that minimize the motion of the black hole on the grid and
which, in the absence of an algorithm to handle a moving excised
region on the numerical grid, would inevitably lead to rapid crash of
the code.  An obvious solution to this problem is to allow for the
motion of the excised region, which, in turn, requires a proper
handling of the regions of the grid that need no longer to be excised,
but rather to be populated with relevant data
(see~\textit{e.g.}~\cite{Calabrese2003:excision-and-summation-by-parts,
  Pretorius05,Sperhake2005a}). While this is technically possible, it is not yet
implemented in our code. Alternatively, it is possible to increase
progressively the size of the excision region while making sure that
all points inside this region stay inside. Clearly, this is a solution
which has only a limited validity and works well only for very small
mass ratios or up until the excised region has grown to be comparable
with the size of the apparent horizon. A third possibility, and which
has proven to be successful in our implementation, is to minimize the
black hole movement by letting the whole grid move towards it. In
practice, this entails a suitable use of the gauge conditions for the
shift based on the knowledge of the position of the apparent horizon.
This idea is not new (see, for instance, ref.~\cite{Anninos94e}) and
amounts to correcting the Gamma-driver shift
condition~\eqref{eq:hyperbolicGammadriver} with an additional term
which reflects a coordinate acceleration towards the black hole. This
is most conveniently done by modeling the changes as those experienced
by a damped harmonic oscillator~\cite{Bruegmann:2003aw}.

More specifically, assuming $x_m$ to be the coordinate position of the
apparent horizon and $x_d$ the desired one so that $x_h\equiv x_m-x_d$
measures the deviation from the optimal position, the contribution to the
shift evolution coming from this correction can be calculated using the
damped harmonic-oscillator equation
\begin{equation}
 T^2\ddot{x}_h + 2Td\dot{x}_h + x_h = 0
\end{equation}
where $T$ is the frequency of the harmonic oscillation and $d/T$ the
decay time for the oscillation. In practice this leads to an acceleration
\begin{equation}
\ddot{x}_h = -\frac{1}{T^2}(2Td\dot{x}_h+x_h),
\end{equation}
which is added to the right-hand-side of the evolution equation for the
shift vector~\eqref{eq:hyperbolicGammadriver}.

The constants $T$ and $d$ need to be chosen accurately in order to
obtain the desired behavior and our experience is that excessively
small values for $T$ or excessively large values for $d$ can lead to
instabilities. Similarly, a too large value for $T$ reduces the effect of
this correction, while too small a value for $d$ can lead to undesired
oscillations in the shift. Overall, we have found that values of $T=10$
and $d=1$ work best in our particular case.  Using this correction, the
position of the apparent horizon can be held fixed on the numerical grid,
which increases the life-time of the simulations from $\approx 230~M$ to
$\approx 1700~M$.

\begin{figure}
 \centering
 \includegraphics[width=8.5cm]{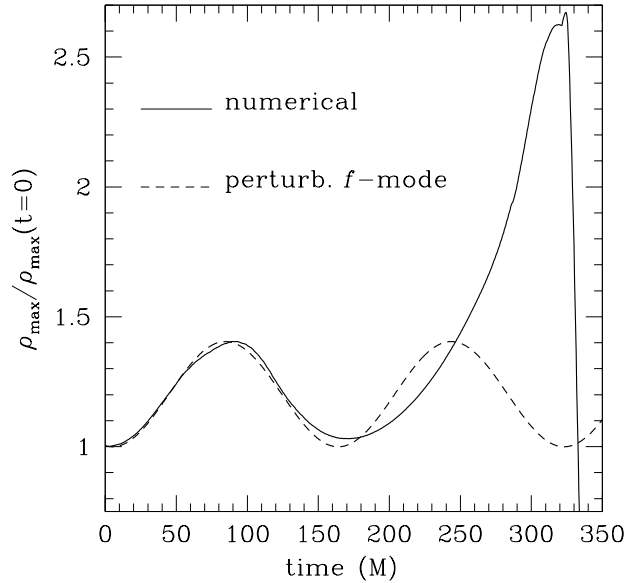}\\
 \caption{Shown is the maximal rest-mass density $\rho$ over time (solid
          line) and as comparison a trigonometric function with the
          frequency of the fundamental mode of the unperturbed star of
          which is $\approx 1.28\mbox{kHz}$.}
 \label{fig:tp_bhns_rho_max}
\end{figure}

\subsection{Dynamics of the Collision}

\begin{figure*}
 \centering
 \vspace{-0.55cm}\includegraphics[width=12cm]{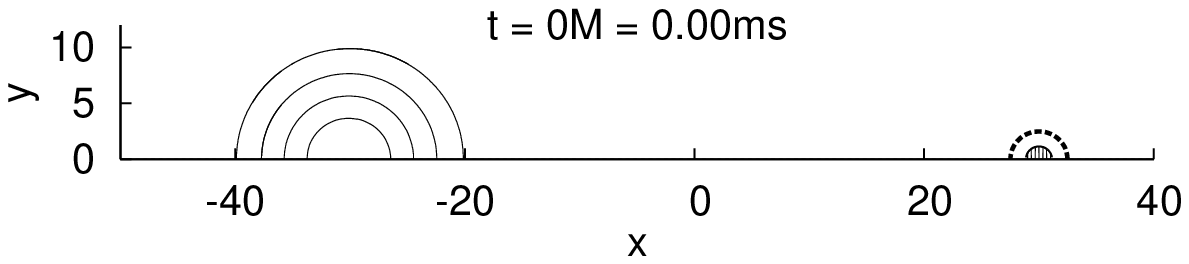}\\
 \vspace{-0.55cm}\includegraphics[width=12cm]{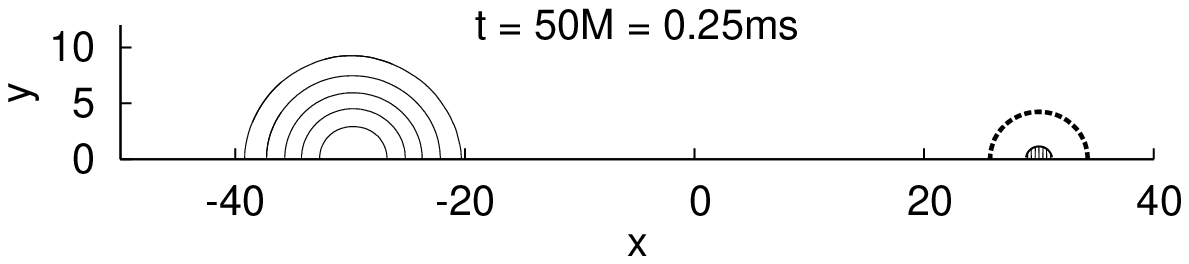}\\
 \vspace{-0.55cm}\includegraphics[width=12cm]{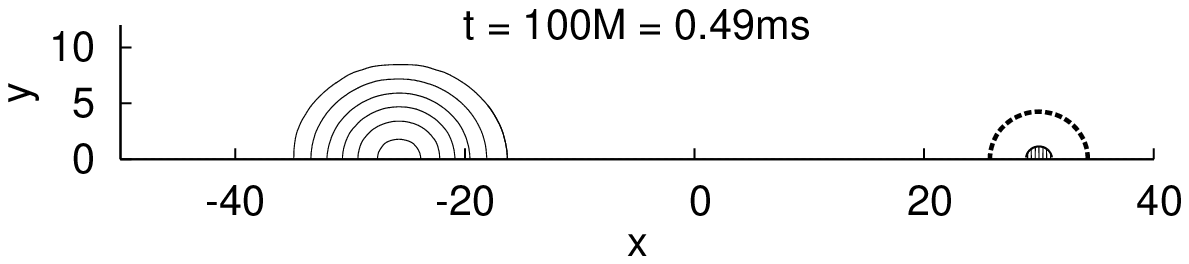}\\
 \vspace{-0.55cm}\includegraphics[width=12cm]{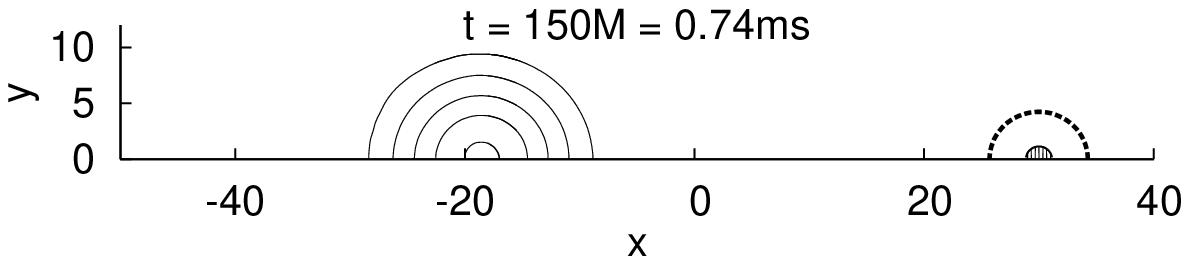}\\
 \vspace{-0.55cm}\includegraphics[width=12cm]{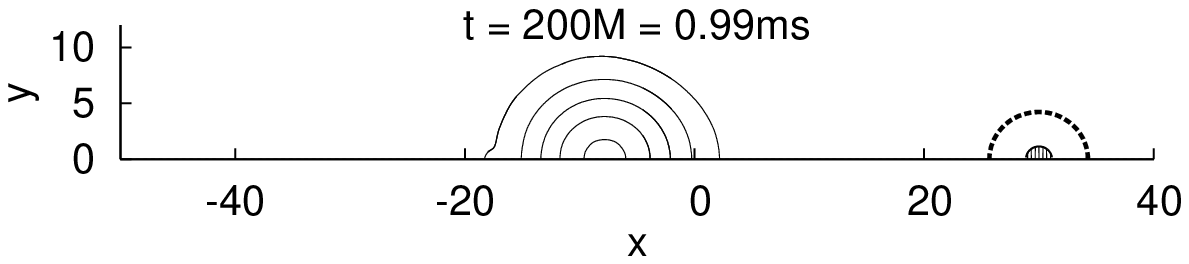}\\
 \vspace{-0.55cm}\includegraphics[width=12cm]{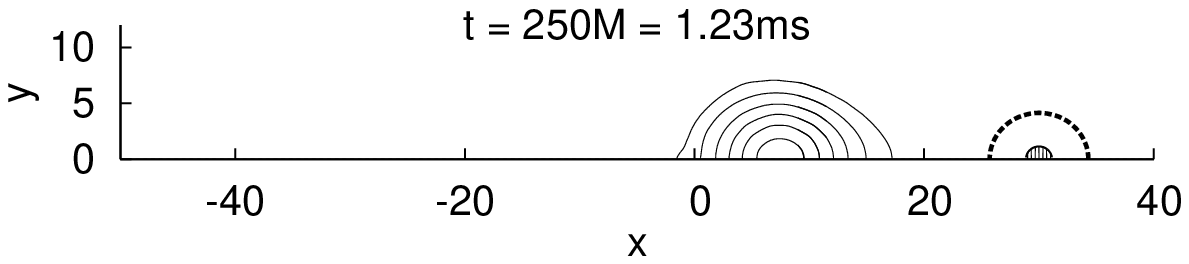}\\
 \vspace{-0.55cm}\includegraphics[width=12cm]{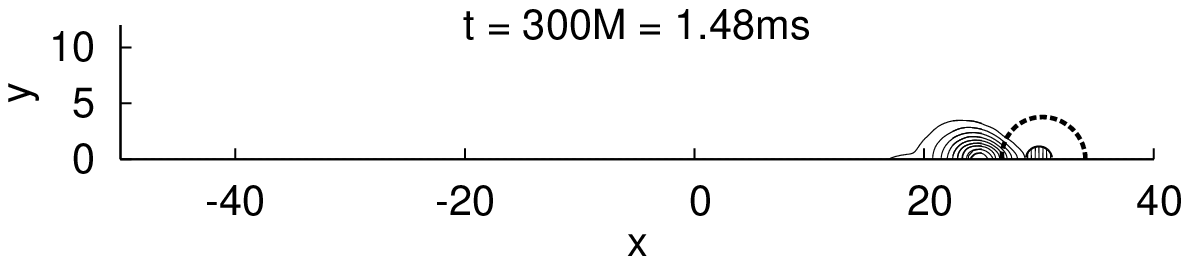}\\
 \vspace{-0.55cm}\includegraphics[width=12cm]{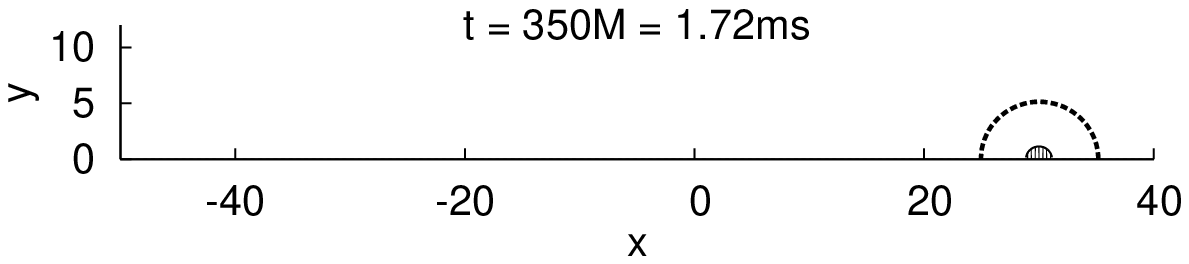}\\
 \caption{Isocontours of the rest-mass density and of the apparent
          horizon surface (dashed line) in the $(x,y)$ plane; the area
          shown as filled refers to the excised region of the
          computational domain. The different panels refer to times from
          $0~M$ to $350~M$ in steps of $50~M\simeq 0.25$~ms.}
 \label{fig:2dsnaps_1}
\end{figure*}

\begin{figure*}
 \centering
 \hspace{-0.5cm}\includegraphics[width=8cm]{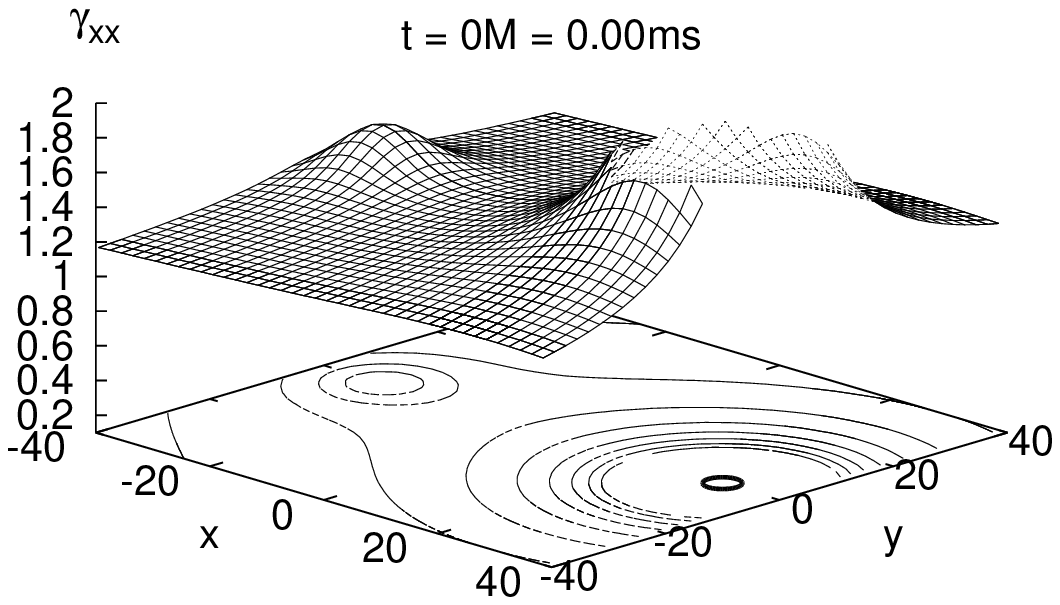}
 \hspace{-0.5cm}\includegraphics[width=8cm]{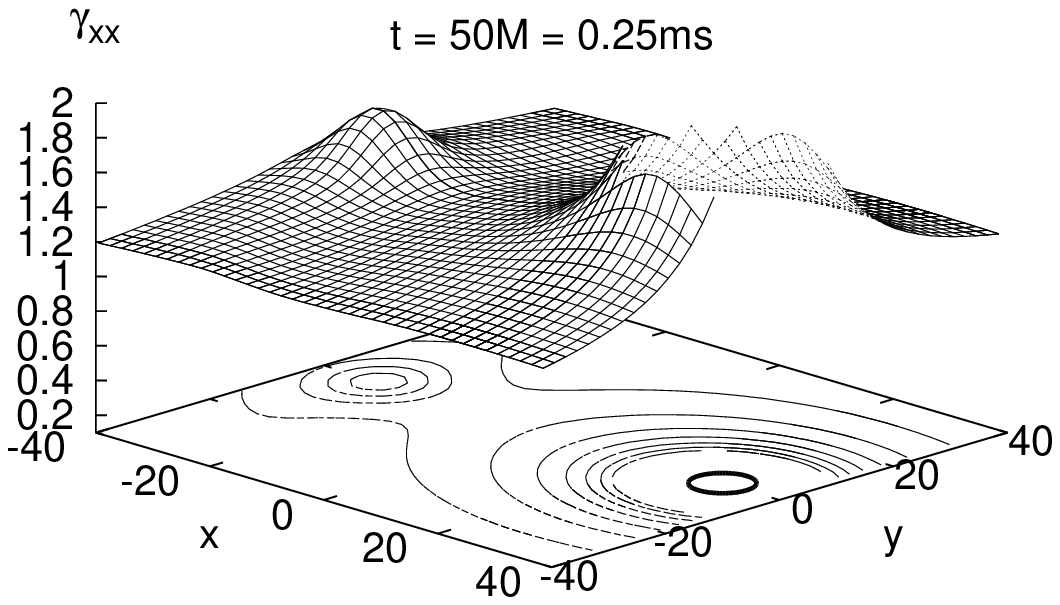}\\
 \vspace{-0.45cm}
 \hspace{-0.5cm}\includegraphics[width=8cm]{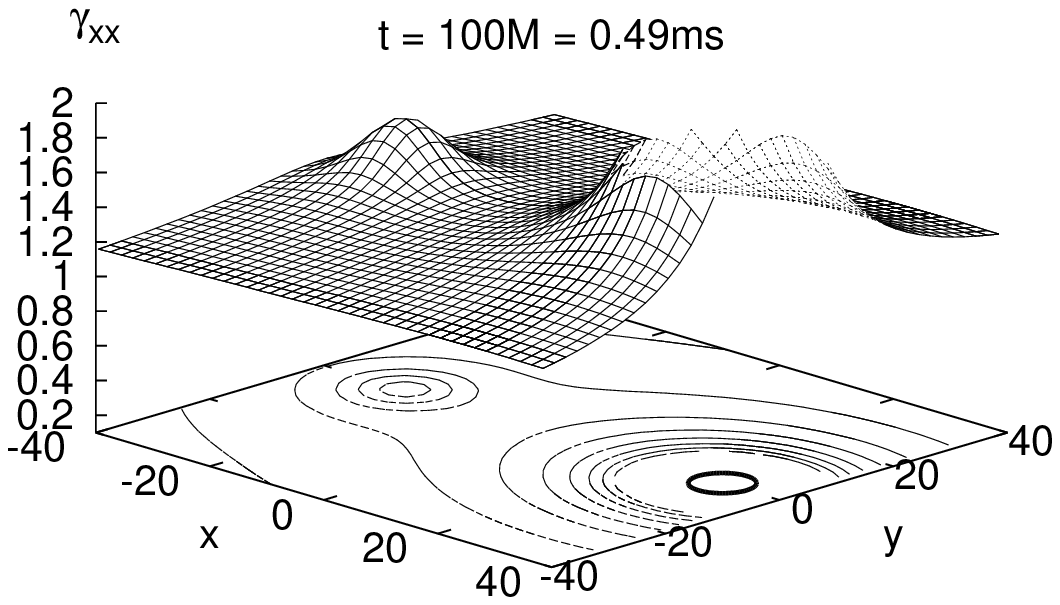}
 \hspace{-0.5cm}\includegraphics[width=8cm]{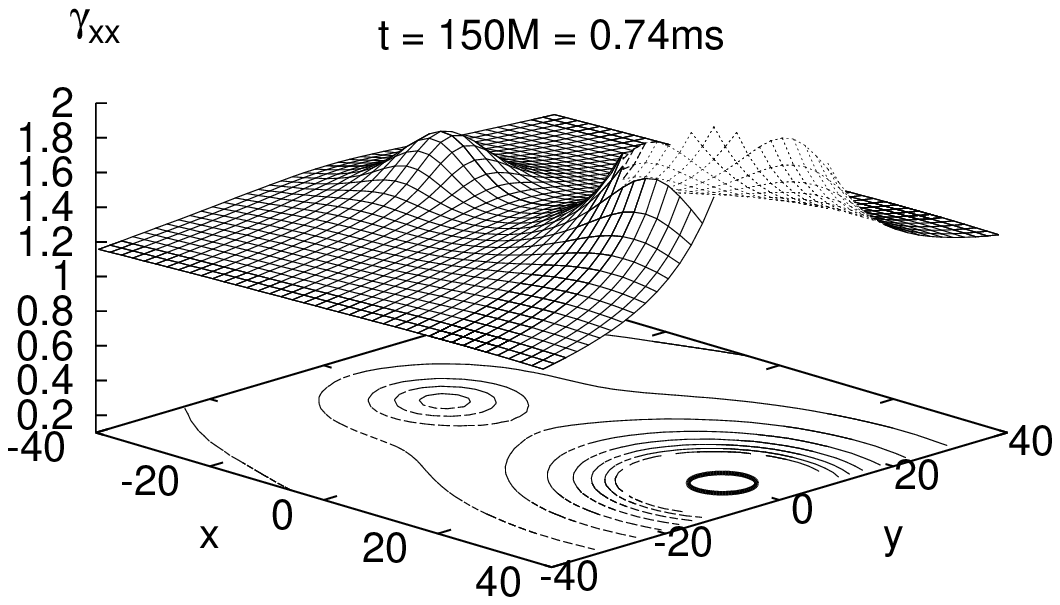}\\
 \vspace{-0.45cm}
 \hspace{-0.5cm}\includegraphics[width=8cm]{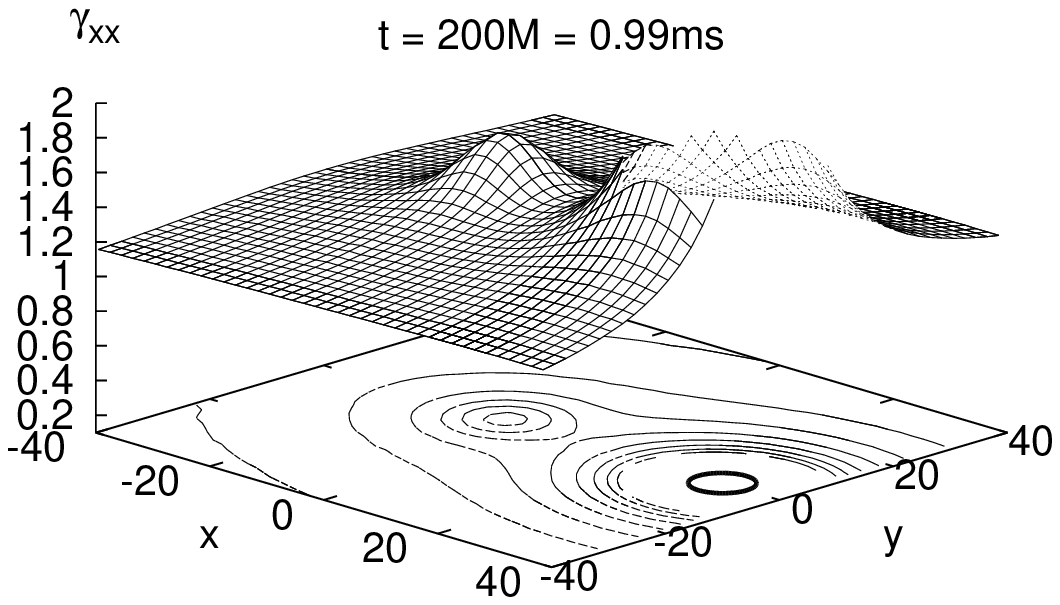}
 \hspace{-0.5cm}\includegraphics[width=8cm]{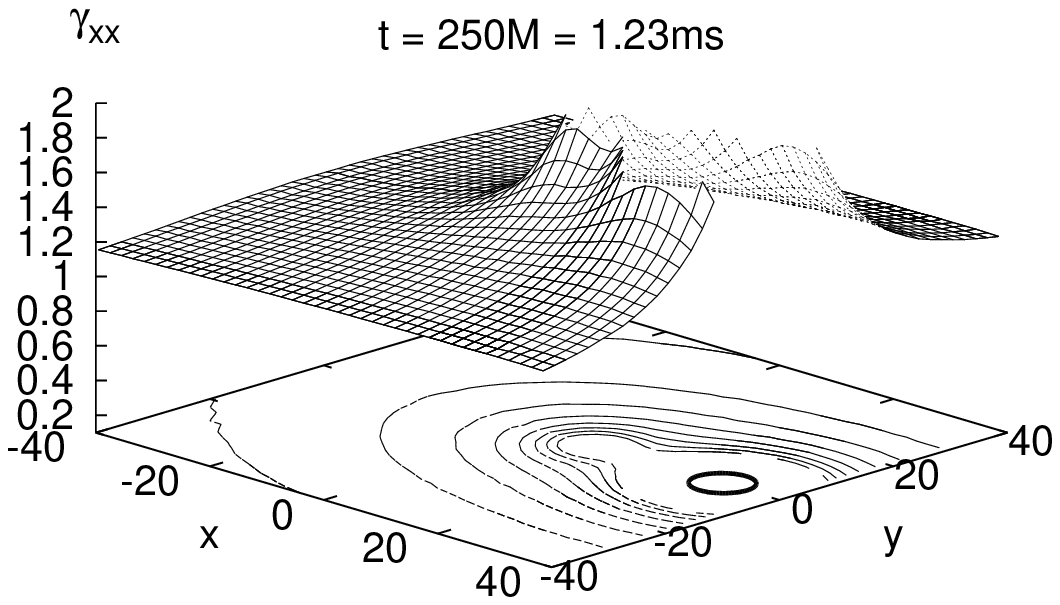}\\
 \vspace{-0.45cm}
 \hspace{-0.5cm}\includegraphics[width=8cm]{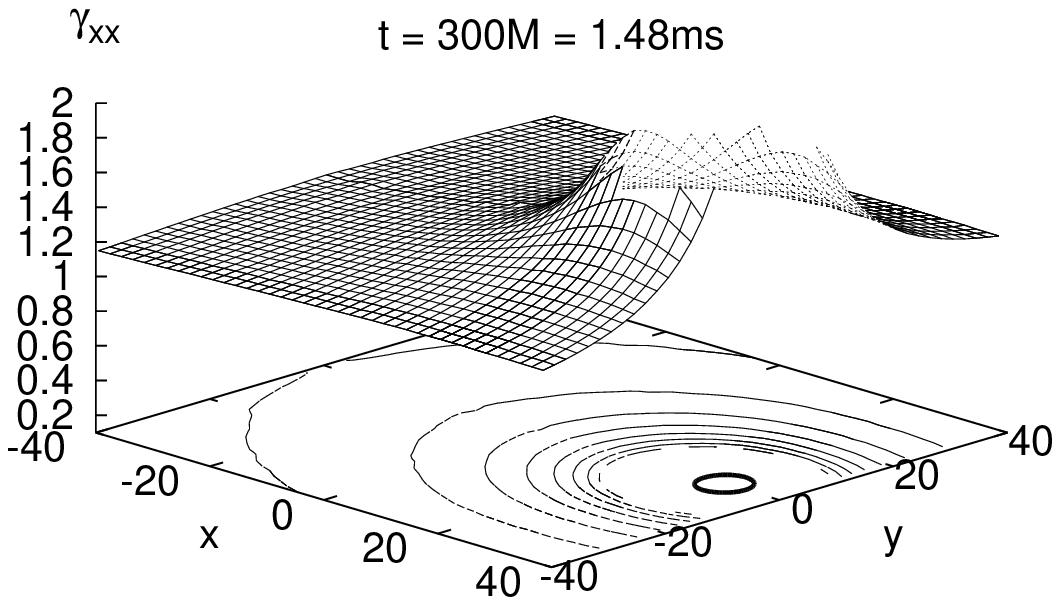}
 \hspace{-0.5cm}\includegraphics[width=8cm]{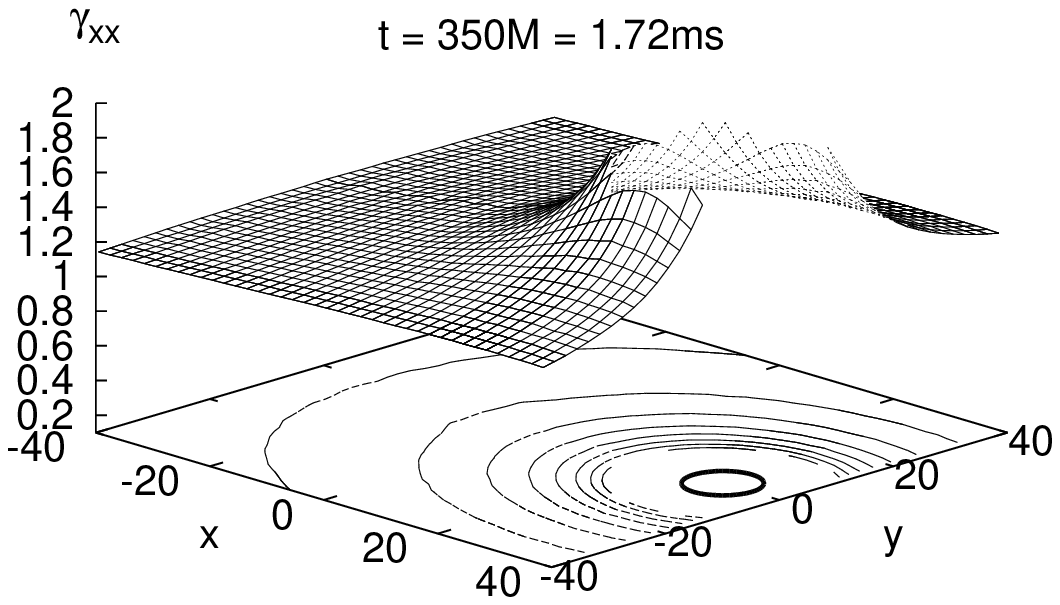}\\
 \caption{Time evolution of a representative metric function ({\it i.e.,}
   $g_{xx}$), in its cross-section onto the $(x,y)$ plane, with the circle
   across the $y=0$ axis showing the position of the apparent horizon. The
   different panels refer to times from $0~M$ to $350~M$ in steps of
   $50~M\simeq 0.25$~ms and we have reported only the solution evaluated
   on the coarsest grid.}
 \label{fig:2dsnaps_2}
\end{figure*}

\begin{figure*}
\centering
\hspace{-.5cm}
\includegraphics[width=9.cm]{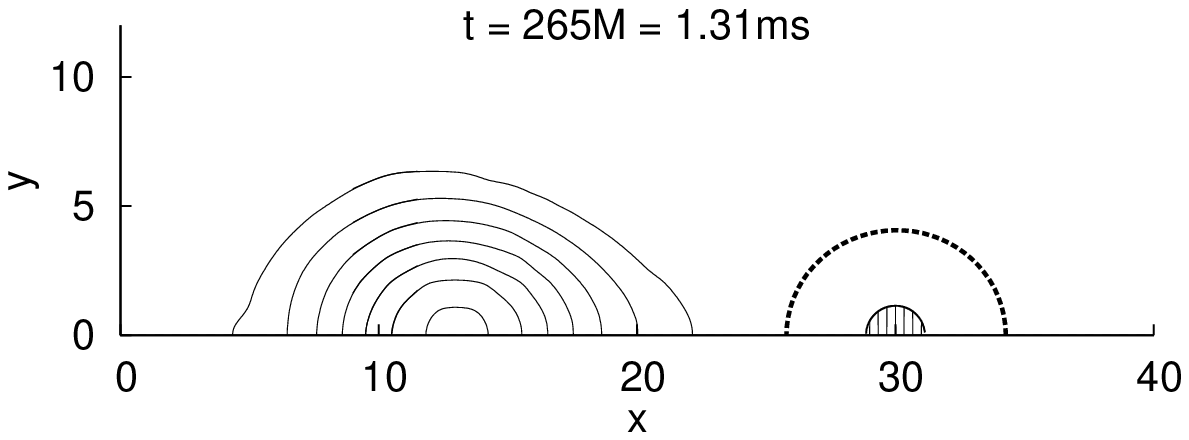}
\includegraphics[width=9.cm]{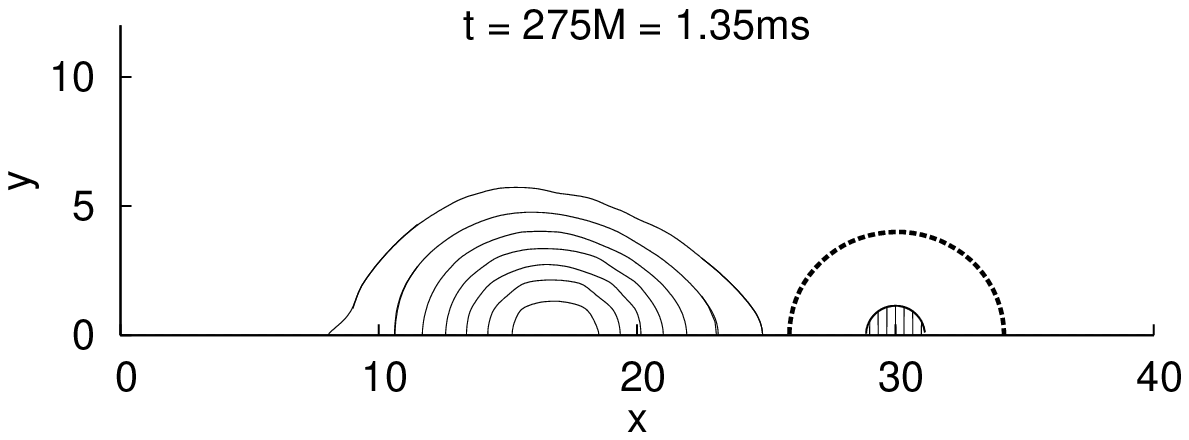}\\
\hspace{-.5cm}
\includegraphics[width=9.cm]{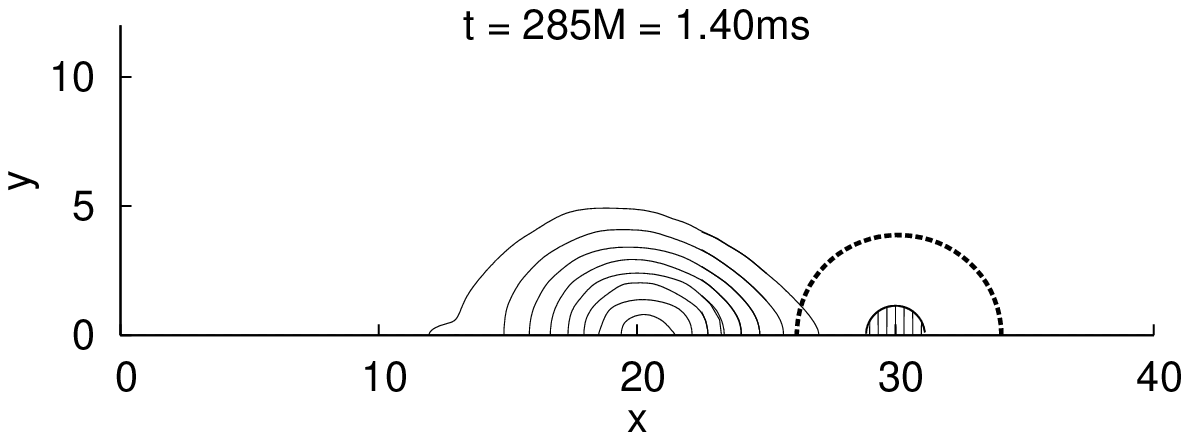}
\includegraphics[width=9.cm]{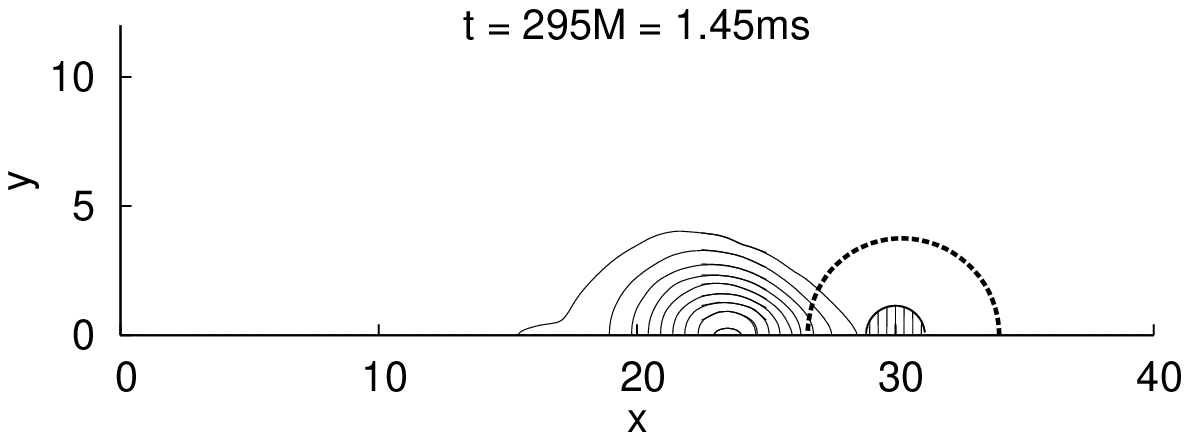}\\
\hspace{-.5cm}
 \includegraphics[width=9.cm]{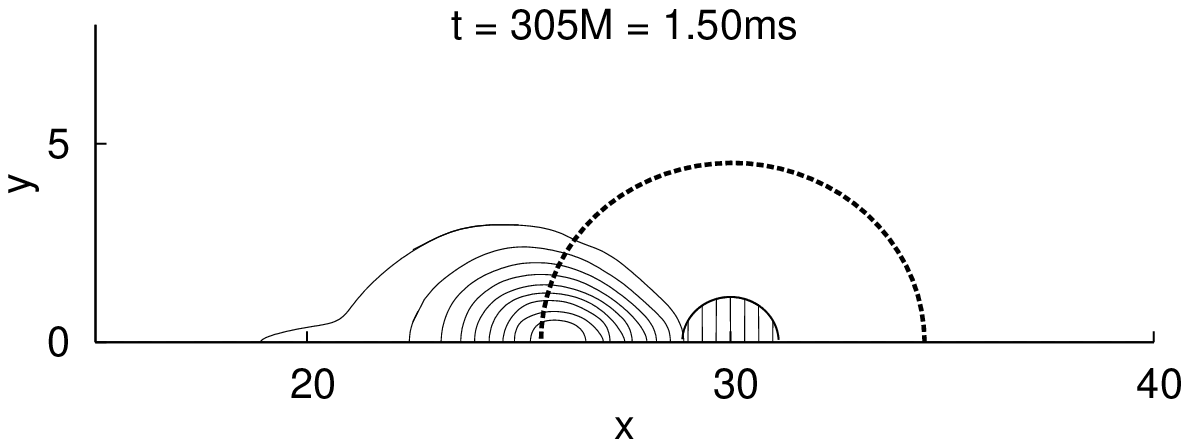}
 \includegraphics[width=9.cm]{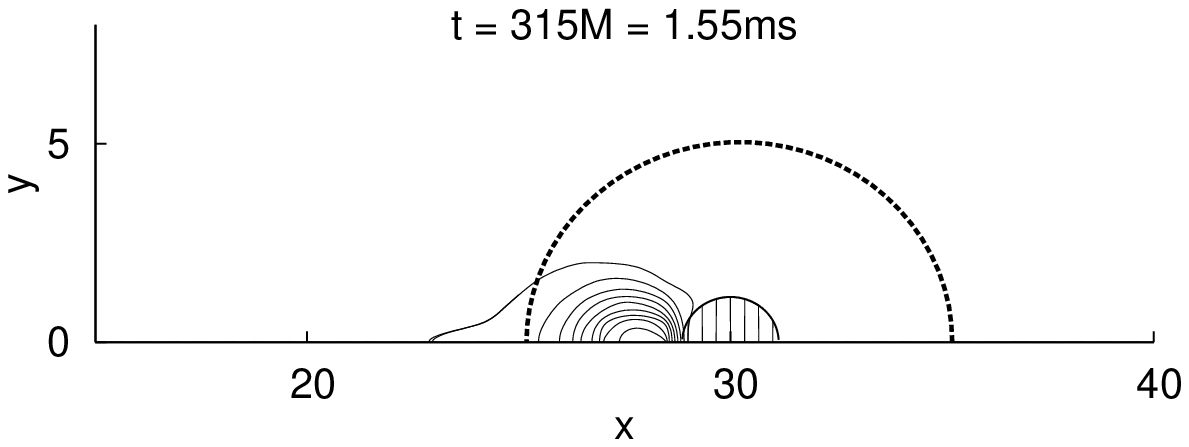}\\
\hspace{-.5cm}
 \includegraphics[width=9.cm]{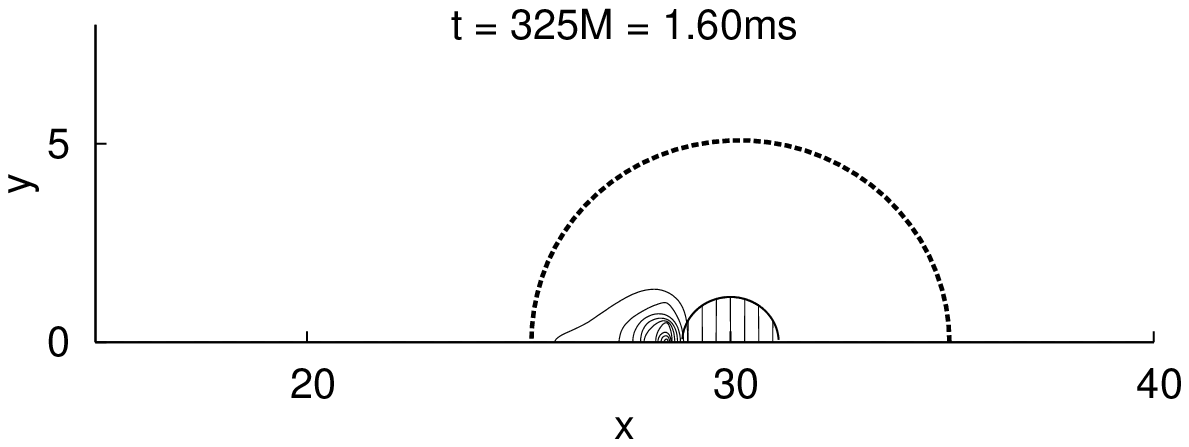}
 \includegraphics[width=9.cm]{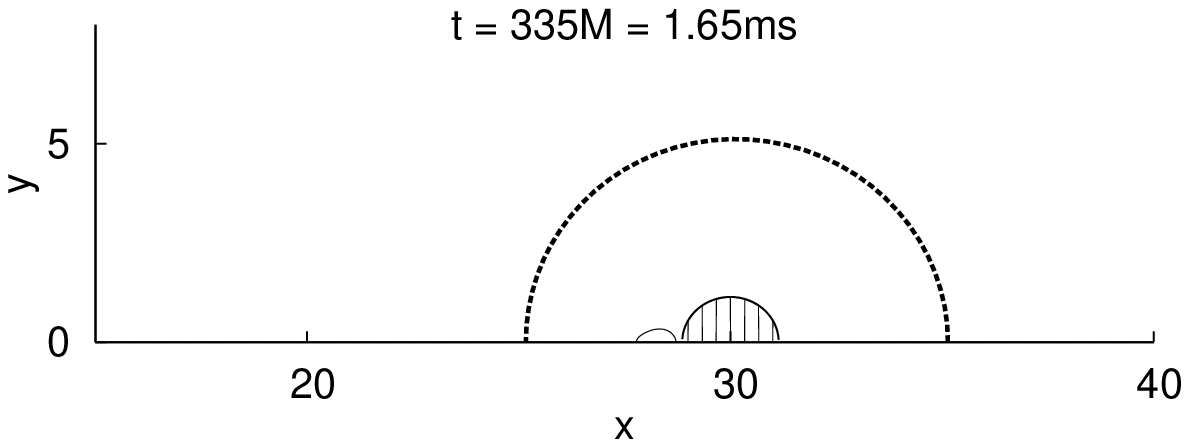}\\
 \caption{Isocontours of the rest-mass density and of the apparent
          horizon surface (dashed line) in the $(x,y)$ plane; the area
          shown as filled refers to the excised region of the
          computational domain. The different panels refer to times from
          $265~M$ to $335~M$ in steps of $10~M$. Note that the scale in
          the first four frames is different (larger) than the one in the
          last four frames.}
 \label{fig:2dsnaps_3}
\end{figure*}

Although our approach is rather generic and could allow for a very
large range of mass ratios and separations between the two compact
objects, hereafter we will concentrate on a prototypical mixed binary
system consisting of a initial guess for the neutron star with a
gravitational mass of
$0.86~M$, proper radius $R= 10.49~M$ ({\it i.e.,} $M/R=0.08$), and a
Schwarzschild black hole with a mass of $5~M$.  The coordinate
separation is $60~M$ [{\it i.e.,} $b=30$ in
  eq.~\eqref{eq:coordinate_transformation}]. The initial velocities
are taken to be zero and the initial data conformally flat and
time-symmetric. The ADM mass, as measured on the compactified,
spectral grid, is $5.78~M$.

As one would expect, the combined perturbations coming from the
truncation error ({\it cf.,} results from the evolution of isolated
stars~\cite{Baiotti04}) and from the black-hole's tidal
field~\footnote{Indeed, we have verified that the tidal field is the
  largest source of perturbation. In particular, we have performed a
  number of simulations with the same spatial resolution but having
  different initial separations $2d$. If the oscillations were
  produced by the truncation error, their amplitude would be almost
  unchanged for different $d$; on the contrary we have found that the
  amplitude scales rather well with $1/d$, which is what one would
  expect from a perturbation induced by a tidal field.}, induce
oscillations dominated by the fundamental mode. These oscillations
appear to be harmonic up until the gravitational influence of the
black hole becomes apparent and therefore changes the nature of the
oscillating mode. In our simulation, the star is not in
equilibrium. This perturbation also produces oscillations like the
ones introduced by the truncation error, but with much bigger
amplitude. This is illustrated in Fig.~\ref{fig:tp_bhns_rho_max} where
we show with a solid the time evolution of the maximum rest-mass
density (which coincides with the rest-mass density at the center of
the star). Indicated with a dashed line is also a trigonometric
function at the frequency of the fundamental ($f$) mode of the
unperturbed star which was measured to be $\approx 1.28\mbox{kHz}$ in
a separate simulation. Note also that the star oscillates for a bit
longer than a full period before the it is tidally disrupted and fully
accreted onto the black hole, which occurs at $\simeq 350~M$. The
central density rises considerably during the accretion process with a
net increase of a factor of about $2.5$. For stars that are in
eccentric orbits with very close periastrons and that are sufficiently
close to the stability threshold, the monotonic increase in central
density produced by the black-hole's tidal field could lead to either
phase transition~\cite{lin_etal_06} or to the collapse to a black
hole. In both cases, which we plan to investigate in subsequent work,
a copious emission of gravitational wave would be produced.

	A self-explanatory description of the matter evolution is
presented in Fig.~\ref{fig:2dsnaps_1} which shows consecutive snapshots
of the stellar motion towards the black hole. Different panels are
separated in time of about $50~M\simeq 0.25$ ms (time progresses from the
top to the bottom of the page) and refer to logarithmically spaced
isocontours of the the rest-mass density in the $(x,y)$ plane. Also shown
with a dashed line is the coordinate shape of the apparent horizon, while
the filled area in its interior refers to the excised region.

\begin{figure}
 \centering
 \centering \includegraphics[width=8.5cm]{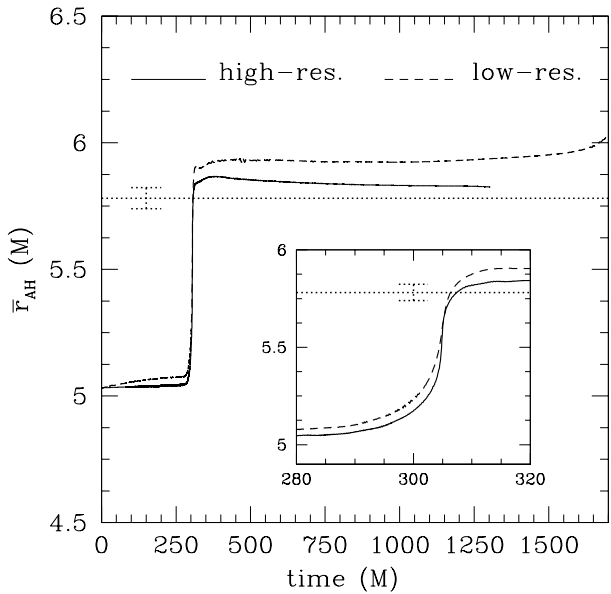}\\
 \caption{Time evolution of the mass of the apparent horizon shown for
          a high-resolution simulation (solid line) and a low-resolution
          one (dashed line). Indicated with a horizontal dotted line is
          the expected ADM mass and the error-bar indicates the error
          introduced by a finite-size domain for the smallest resolution
          used. The inset shows the evolution around the time of the
          accretion of the neutron star onto the black hole.}
 \label{fig:ah_mass}
\end{figure}

\begin{figure}
 \centering
 \includegraphics[width=8.5cm]{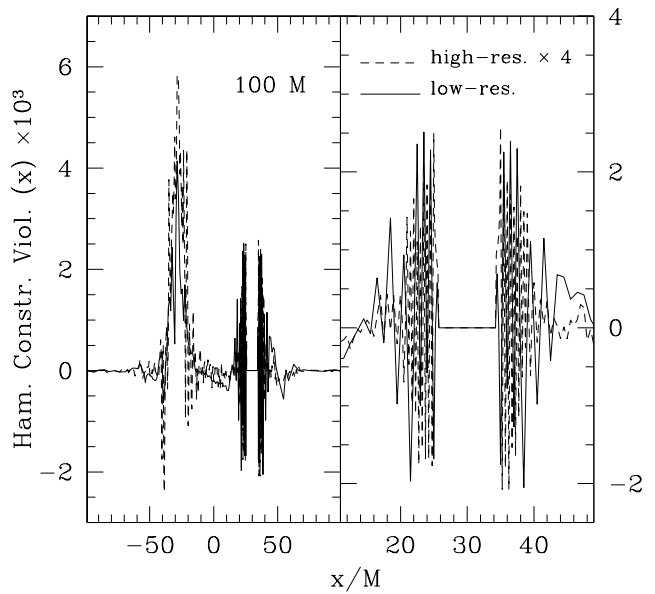}\\
 \caption{1D slice of the Hamiltonian constraint violation along the
   $x$-axis at a representative time $t=100\,M$. Each panel reports
   two curves indicating respectively the violations computed with a
   low-resolution grid (solid line) and with a high-resolution one
   (dashed line) scaled by a factor of 4 for the expected second-order
   accuracy. While the left panel shows the portion of the grid
   containing the two compact objects, the right one focuses on the
   violation around the black hole.}
 \label{fig:tp_bhns_evol_ham}
\end{figure}

Note that the stellar model is initially spherically symmetric and that
in about a sonic crossing time it starts to move towards the black
hole. As it does so, it also experiences an increasingly strong tidal
field which leads to a flattening of the star which then assumes the
typical olive-shaped profile. The accretion process takes place between
$t\simeq 280~M$ and $t\simeq 340~M$, so that by $t\simeq 350~M$
essentially no matter is left outside the apparent horizon. Note also
that the evolution proceeds well after this time and indeed an accurate
evolution with moderate growth of the constraint violation is possible up
to a time $t \sim 1700~M$, {\it i.e.,} approximately $1300~M$ past when
the last stellar fluid-element has been accreted onto the black hole (see
also Figs.~\ref{fig:ah_mass} and \ref{fig:tp_bhns_evol_ham})~\footnote{We
recall that strictly speaking the spacetime is not pure vacuum also after
the last stellar fluid-element. This is because of the presence of the
very tenuous atmosphere introduced to use the high-resolution
shock-capturing techniques and which does not influence perceptibly the
overall dynamics~\cite{Baiotti03a,Baiotti04}}. We consider this a
significant achievement in numerical simulations involving non-vacuum and
non-stationary black hole spacetimes. To underline how this represents
remarkable improvement with respect to the present state-of-the-art
calculations, we note that in our recent investigations of the collapse
of a rotating star to a Kerr black hole~\cite{Baiotti04}, we were able to
prolong the evolution of $\sim 30~M$ past the appearance of the apparent
horizon and the introduction of an excised region and only $\sim 10~M$
after the star was fully accreted onto the excised region.

Fig.~\ref{fig:2dsnaps_2} shows the corresponding evolution across the
$(x,y)$ plane of a representative metric function, {\it i.e.,} $g_{xx}$,
during the collision and accretion. The solution shown refers to the
coarsest grid only and the circle across the $y=0$ axis shows the
position of the apparent horizon. Also in this case the evolution
presented in the different panels is self-explanatory and worth
underlining is only the regularity of the solution during the accretion
process and the rapid recovery of the solution for an isolated black hole
after $t\simeq 300~M$.

Fig.~\ref{fig:2dsnaps_3} offers a closer look of the final stages of the
collision starting from $t=265~M$ and with a progression of $10~M$ (time
progresses from the upper-left corner to the lower-right one). Note the
progressive flattening of the neutron star produced by the tidal field
and the long ``tail'' resulting from the strong rarefaction and produced
by the rapid accretion from $t \simeq 295~M$ to $t\simeq 325~M$. Note
also that the matter dynamics is accurately described well within the
apparent horizon (dashed thick line in Fig.~\ref{fig:2dsnaps_2}) and that
the motion into the excised region (shaded area in
Fig.~\ref{fig:2dsnaps_2}) takes place very smoothly without the
appearance of irregularities in the flow. This is the result of
self-consistent treatment of the flow across the excision region and
which makes full use of the characteristic information~\cite{Hawke04}.

The sequences shown in Figs.~\ref{fig:2dsnaps_1}-~\ref{fig:2dsnaps_3}
are also useful to appreciate the effectiveness of the improved gauge
conditions. Note, in fact, how the position of the black hole does not
change perceptibly despite the binary system has a mass ratio of
$\simeq 5.8$. When looked at carefully, the apparent horizon increases
its coordinate dimensions during the early stages of the evolution as
it does also as a result of the accreted stellar mass. Clearly, the
first of these growths is just a coordinate effect resulting from our
gauge choice which moves grid points outwards from the apparent
horizon and leads to its coordinate growth. Indeed, when looked at in
terms of its proper dimensions, the apparent horizon does not show a
significant increase during the early stages of the evolution and its
growth is essentially confined to the time during which matter is
accreting onto it. This is shown in Fig.~\ref{fig:ah_mass} which
presents the time evolution of the horizon mass for both a low and a
high-resolution simulation (dashed and continuous curves,
respectively) and compares it with the expected ADM mass ({\it i.e.,}
as measured on the Cartesian grid and indicated with a dotted line)
and the error bar resulting from the use of a computational domain of
finite extent.

A number of comments are needed here. Firstly, it should be noted that
apparent horizon mass is not constant before the merger but shows a small,
constant drift to higher masses. This is clearly the result of a large
truncation error, especially near the apparent horizon, and is
considerably reduced in the high-resolution run. Secondly, it is apparent
that the increase in the horizon mass is essentially limited to the time
interval during which the neutron star is accreted onto the black hole,
that both measures are compatible with the expected error-bar and that as
the resolution is increased the error in the expected mass-growth before
merger is also progressively reduced. Finally, note that after the merger
the mass of the apparent horizon does not remain constant but rather
decreases slightly, so that by the time the simulation is completed, the
black hole has lost about $0.7\%$ of its mass. While a mass-loss is in
principle expected as a result of the emission of gravitational radiation
from the oscillating black hole, in Section~\ref{gwe} we will show that
indeed the latter is much smaller and hence the secular mass loss in the
post-merger evolution is small and essentially due to the truncation
error.

\begin{figure}
 \centering \includegraphics[width=8.5cm]{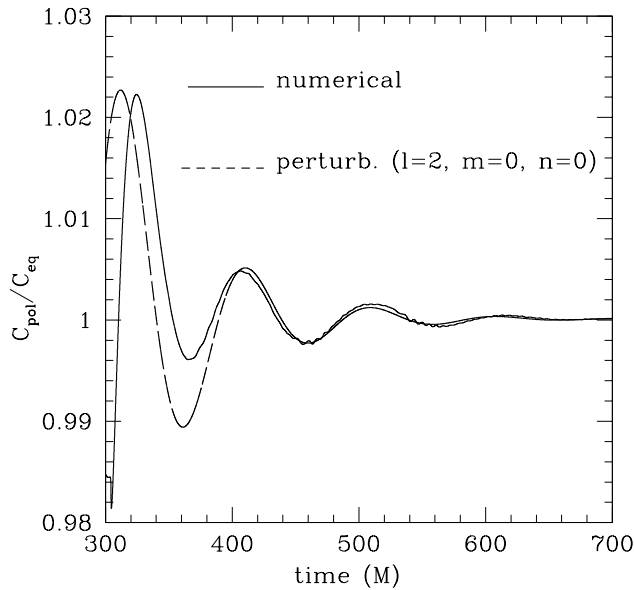}\\
 \caption{Time evolution of the ratio between the polar and equatorial
   proper circumferences of the apparent horizon soon after the collision
   (solid line). Shown for
   a comparison with a dashed line is the QNM oscillation of a black hole
   with mass $\simeq 5.86~M$, fundamental $\ell=2$ frequency of $f\approx
   2.06\mbox{kHz}$ and damping time of $0.32~{\rm ms}$.}
 \label{fig:tp_bhns_AH_radius_ratio}
\end{figure}

Besides the growth in area, the black hole is also expected to
experience quasi-normal mode (QNM) oscillations as a response to the
collision. We have carefully looked at whether these oscillations are
detectable and we show with a solid line in
Fig.~\ref{fig:tp_bhns_AH_radius_ratio} the time evolution of the ratio
between the polar and equatorial proper circumferences soon after the
collision has taken place. In order to assess whether this damped
oscillation does correspond with the expected QNM we compare this
behavior with the one expected from perturbative analyzes which
predict the frequency of the $\ell=2$ fundamental ($n=0$) QNM to be
~\cite{Leaver85,Kokkotas99a}
\begin{equation}
 \label{eq:bhns_bh_qnm_f}
f \simeq 0.374 \times 2\pi(5.14\mbox{kHz})\times (M_\odot/M) \ .
\end{equation}
or equivalently $f=2.06\mbox{kHz}$ for a black hole mass of $\simeq
5.86~M$ and an expected damping time of $0.32~{\rm ms}$ for the same
mass. The corresponding QNM behavior is shown as a dashed line in
Fig.~\ref{fig:tp_bhns_AH_radius_ratio} for the high-resolution
simulation. The rather good agreement is an
important confirmation that the oscillation detected is not a
numerical by-product but embodies important physical
information. Perhaps even more impressive than the good matching with
the perturbative results is the consideration that the QNM
oscillations shown in Fig.~\ref{fig:tp_bhns_AH_radius_ratio} have
indeed a very small amplitude of less than $2\%$; the ability of our
code to detect it is a confirmation of the overall accuracy achieved
but also an indication that high-resolution near the black hole
apparent horizon is crucial for a consistent modeling of the
black-hole dynamics.

We conclude this Section by showing in Fig.~\ref{fig:tp_bhns_evol_ham}
a one-dimensional (1D) slice of the Hamiltonian constraint violation
along the $x$-axis at a representative time $t=100\,M$. Each panel in
the Figure reports two curves at two different resolutions, the solid
line referring to a low-resolution grid and the dashed one to a
high-resolution grid rescaled by a factor of 4 to highlight the
second-order convergence. Furthermore, while the left panel shows the
portion of the grid comprising the two compact objects, the right one
offers a closer look at the violation in the vicinity of the black
hole. A couple of aspects of Fig.~\ref{fig:tp_bhns_evol_ham} are worth
highlighting. Firstly, the rapid variation of the constraint violation
shown in the Figure is due to the slightly different truncation errors
in the regions where two refinement levels overlap and which is
magnified when the second spatial derivatives are calculated in the
calculation of the Hamiltonian constraint. Secondly, the rate of
convergence is different in different parts of the grid, being
somewhat less than second-order over the regions covering the matter
sources and which are probably not sufficiently resolved. Finally, the
convergence rate is higher in those regions of the domain where the
fields do not vary rapidly (we recall that
Fig.~~\ref{fig:tp_bhns_evol_ham} shows only a small portion of the
numerical grid which extends from $x\sim -200\,M$ to $x\sim 200\,M$).

Overall, the evolution of more global quantities, such as the 2-norm
of the Hamiltonian constraint violation show that the norm increases
by roughly three orders of magnitude soon after the simulations
start. This simply indicates that the errors in the determination of
the initial data are much smaller than the ones introduced by the
evolution of these data. At the time of merger ({\it i.e.}, at
$t\approx 300~M$), the violation increases rapidly by roughly an order
of magnitude (and can be locally even larger) as a result of the
matter crossing many mesh refinement boundaries and finally the
excision boundary. Once the collision has taken place and all of the
matter has been accreted onto the black hole, the constraint violation
settles to a constant but somewhat smaller value than the one before
the merger since no matter is left in the computational domain and
this suppresses the contribution to the violation of the Hamiltonian
constraint coming from the matter terms [{\it cf.} eq.(2.6)
  in~\cite{Baiotti04}]. Also, the code maintains second-order
convergence up until the errors coming from the excision boundary
spoil the convergence and thus increase the truncation error. From
shortly after the merger up to $t\approx 1700~M$ the violation is
roughly constant, until a rapid instability appears which grows
exponentially and leads to a code crash. It is still unclear what is
the precise source of this instability, although it is apparent that
the boundary conditions we use are not constraint-preserving and may
be behind the onset of the instability. Work is now in progress to
improve the outer boundary conditions making use of the numerous
results which have been obtained recently in the literature about
this, \textit{e.g.}
refs.~\cite{Abrahams97a,Rezzolla99a,Calabrese:2002xy,Gundlach:2004jp,
  Szilagyi02a}.

\section{Gravitational-wave emission}
\label{gwe}

Several different methods can be used to extract gravitational waves from
a numerical simulation. Here we use a gauge-invariant approach, in which
the numerical spacetime is matched to perturbations of a Schwarzschild
black hole (see refs.~\cite{Rupright98,Camarda97c,Allen98a}).  Because
our numerical grid for the evolution does not extent to spacial infinity,
we have to measure the gravitational wave content at a finite distance
from the final black hole. In practice we calculate the gauge-invariant
metric perturbations $Q^{\rm{(odd)}}_{lm}$ and
$\Psi^{\rm{(even)}}_{lm}$~\cite{Moncrief74} as observed on spheres with
constant radial coordinate distance from the final black hole. $l$ and
$m$ are the indices of the angular decomposition, of which we compute up
to $l=5$ and restrict to $m$ to $0$, because modes with $m\ne0$ are
essentially zero due to the axisymmetry of the merger.

Using the odd and even-parity perturbations defined as
$Q^{\times}_{lm}\equiv\lambda Q^{\rm (odd)}_{lm}$ and
$Q^+_{lm}\equiv\lambda \Psi^{\rm (even)}_{lm}$, where
$\lambda\equiv\sqrt{2(l+2)!/(l-2)!}$, we calculate the transverse
traceless gravitational wave amplitudes in the two polarizations
$h_+$ and $h_{\times}$ as
\begin{equation}
 h_+-\mbox{i}h_{\times}=\frac{1}{2r}\sum_{l,m}
   \left(Q^+_{lm} - \mbox{i}\int_{-\infty}^t Q^{\times}_{lm}(t')dt' \right)\,
   _{-2}Y^{lm},
\end{equation}
where $_{-2}Y^{lm}$ is the $s=-2$ spin-weighted spherical harmonic. 

The position of such ``observers'' is arbitrary so long as they are in a
region of spacetime which is reasonably close to that of a Schwarzschild
spacetime. If the distance to the merging binary is chosen to be too
small, the measured perturbations do not contain the same information on
the gravitational waves as those in the far zone. One way to test if the
position of the observers is chosen far enough away from the final black
hole is to extract the perturbations at various different radii. In the
far-field region these perturbations have to overlap when plotted in
retarded time. In the results presented below, we checked that this is
indeed the case.

\begin{figure}
 \centering
 \includegraphics[width=8.5cm]{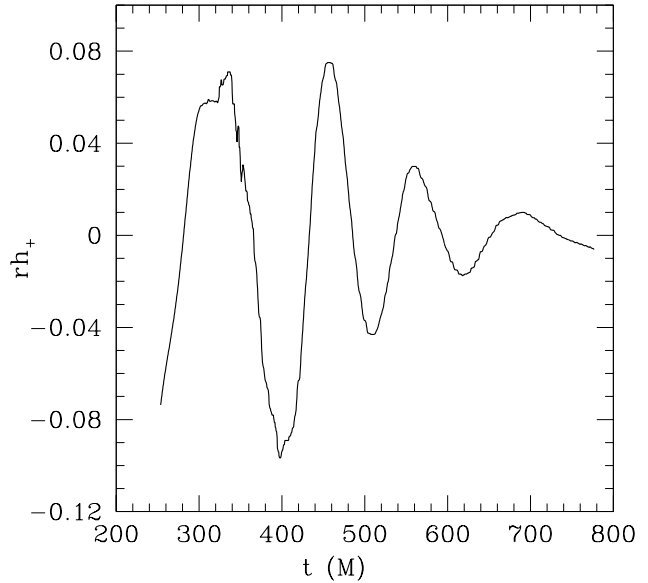}\\
 \caption{$+$ component of the gravitational wave emission from the
   merging binary system, extracted at a radius of $70~M$ from the
   final black hole. The $\times$ component is identically zero.}
 \label{fig:hplus}
\end{figure}

In Figure~\ref{fig:hplus} we show $h_+$ at a (coordinate) distance of
$70~M$ from the final black hole (waveforms extracted from nearby
observers overlap in retarded time confirming that the information is
taken in the wave-zone). $h_{\times}$ is essentially zero because of
the symmetries of the head-on merger.  The emitted power computed as
the total energy lost due to gravitational waves
\begin{equation}
 \frac{\mbox{d}E}{\mbox{d}t} = \frac{1}{32\pi}
   \left(\left|\frac{\mbox{d}Q^+_{lm}}{\mbox{d}t}\right|^2+
         \left|Q^{\times}_{lm}\right|^2\right),
\end{equation}
is $E\approx7.8\times10^{-4}(M/M_\odot)$, which corresponds to $\approx
0.013\%$ of the mass of the binary system. This value strongly depends on
the time span used for the integration and should be taken only as
order-of-magnitude estimate. Overall, this is much smaller than the
$\approx 0.1\%$ obtained in the Newtonian simulations mentioned in
Section~\ref{sec:introduction} which, however, refer to orbiting
configurations and are expected to be more efficient emitters.

\section{Conclusion}
\label{conclusion}

We have discussed the first simulations of a head-on collision of a neutron
star and a black hole of comparable mass in full General Relativity.
This also included the extraction of the gravitational wave signal produced
as well as an order-of-magnitude estimate of the emitted energy.

In this paper, a new approach for obtaining initial data for a mixed
binary system has been discussed, which also has been presented in more
detail in~\cite{Loeffler05}.  This method involves solving the elliptic
equation for the Hamiltonian constraint in the York-Lichnerowicz
conformal decomposition. The currently used solver uses spectral methods
with adapted coordinates on one domain.  These initial data are evaluated
onto a Cartesian grid structure including fixed mesh refinement. The
evolutions are then performed through the full solution of the Einstein
equations combined with an accurate solution of the relativistic
hydrodynamics equations via high-resolution shock-capturing scheme
techniques. Unlike other relativistic studies of such systems, no
limitation is set for the mass ratio between the black hole and the
neutron star, nor on the position of the black hole, whose apparent
horizon is entirely contained within the computational domain.
Concentrating on a prototypical binary system with mass ratio $\sim 6$,
we find that although a tidal disruption is evident, the neutron star is
accreted promptly and entirely into the black hole.

Two important changes with respect to the evolutions discussed in
refs.~\cite{Hawke04,Baiotti04} have been presented, which have been
important for long-term stable evolutions.  The first of these changes is
the use of artificial dissipation, especially to attenuate problems near
the excision boundary.  The second change is related to special gauge
conditions to minimize the motion of the black hole on the numerical
grid. This is necessary because we have not yet implemented the
possibility of a moving excision region.

We believe that these problems of the excision boundary can be solved in
the near future. One reason for this belief is the presence of analytical
studies of such boundaries, \textit{e.g.} based on summation by parts
techniques as in~\cite{Calabrese2003:excision-and-summation-by-parts},
which show much better behavior near the excision boundary.  Another
reason are upcoming codes using multipatch techniques in which it is
natural to align excision boundaries to grid boundaries and which show
extraordinary stability near the excision
region~\cite{Thornburg2004:multipatch-BH-excision_nourl}.  In addition,
evolutions even without excision seem to be possible, provided a very
high resolution and dissipation inside the apparent
horizon~\cite{Baiotti06}. However, even with the present difficulties
encountered in performing long-term stable and accurate calculations of a
spacetime including an excised region, we were able to successfully carry
out evolutions up to $\sim 1700~M$. This is about $\sim 1400~M$ after the
collision has taken place and provides a sufficient time interval to
extract the gravitational-wave signal produced.

We plan to extend the initial data to non-zero momenta in the
future. This would allow for orbiting configurations, which should be no
particular problem for the evolution techniques used. Other improvements,
like the use of a more realistic EOS or the inclusion of magnetic fields
are also desirable and will be focus of future work. Finally, it is worth
mentioning that the techniques described in this paper can in principle
also be applied to the creation of initial data for a binary system of
two neutron stars and its evolution. Comparing the differences in the
dynamics of the merger of black-hole binaries, mixed binaries and
neutron-star binaries, and in particular the different efficiencies in
the gravitational-wave emission, would also be very interesting and the
subject of future work.

\acknowledgments

F.~L\"offler is a VESF fellow of the European Gravitational Observatory
(EGO). Support to this research comes also through the SFB-TR7
``Gravitationswellenastronomie'' of the DFG.

\bibliographystyle{apsrev-nourl}
\bibliography{bhns_headon}

\begin{thebibliography}{72}
\expandafter\ifx\csname natexlab\endcsname\relax\def\natexlab#1{#1}\fi
\expandafter\ifx\csname bibnamefont\endcsname\relax
  \def\bibnamefont#1{#1}\fi
\expandafter\ifx\csname bibfnamefont\endcsname\relax
  \def\bibfnamefont#1{#1}\fi
\expandafter\ifx\csname citenamefont\endcsname\relax
  \def\citenamefont#1{#1}\fi
\expandafter\ifx\csname url\endcsname\relax
  \def\url#1{\texttt{#1}}\fi
\expandafter\ifx\csname urlprefix\endcsname\relax\def\urlprefix{URL }\fi
\providecommand{\bibinfo}[2]{#2}
\providecommand{\eprint}[2][]{\url{#2}}

\bibitem[{\citenamefont{Belczynski et~al.}(2002)\citenamefont{Belczynski,
  Kalogera, and Bulik}}]{Belczynski:2001uc}
\bibinfo{author}{\bibfnamefont{K.}~\bibnamefont{Belczynski}},
  \bibinfo{author}{\bibfnamefont{V.}~\bibnamefont{Kalogera}}, \bibnamefont{and}
  \bibinfo{author}{\bibfnamefont{T.}~\bibnamefont{Bulik}},
  \bibinfo{journal}{Astrophys. J.} \textbf{\bibinfo{volume}{572}},
  \bibinfo{pages}{407} (\bibinfo{year}{2002}).

\bibitem[{\citenamefont{Kalogera et~al.}(2004)\citenamefont{Kalogera, Kim,
  Lorimer, Burgay, D'Amico, Possenti, Manchester, Lyne, Joshi, McLaughlin
  et~al.}}]{Kalogera04}
\bibinfo{author}{\bibfnamefont{V.}~\bibnamefont{Kalogera}},
  \bibinfo{author}{\bibfnamefont{C.}~\bibnamefont{Kim}},
  \bibinfo{author}{\bibfnamefont{D.~R.} \bibnamefont{Lorimer}},
  \bibinfo{author}{\bibfnamefont{M.}~\bibnamefont{Burgay}},
  \bibinfo{author}{\bibfnamefont{N.}~\bibnamefont{D'Amico}},
  \bibinfo{author}{\bibfnamefont{A.}~\bibnamefont{Possenti}},
  \bibinfo{author}{\bibfnamefont{R.~N.} \bibnamefont{Manchester}},
  \bibinfo{author}{\bibfnamefont{A.~G.} \bibnamefont{Lyne}},
  \bibinfo{author}{\bibfnamefont{B.~C.} \bibnamefont{Joshi}},
  \bibinfo{author}{\bibfnamefont{M.~A.} \bibnamefont{McLaughlin}},
  \bibnamefont{et~al.}, \bibinfo{journal}{Astrophys. J. Lett.}
  \textbf{\bibinfo{volume}{601}}, \bibinfo{pages}{L179} (\bibinfo{year}{2004}).

\bibitem[{\citenamefont{Bethe and Brown}(1998)}]{Bethe1998}
\bibinfo{author}{\bibfnamefont{H.}~\bibnamefont{Bethe}} \bibnamefont{and}
  \bibinfo{author}{\bibfnamefont{G.}~\bibnamefont{Brown}},
  \bibinfo{journal}{Astrophys. J.} \textbf{\bibinfo{volume}{506}},
  \bibinfo{pages}{780} (\bibinfo{year}{1998}).

\bibitem[{\citenamefont{{Bejger} et~al.}(2005)\citenamefont{{Bejger},
  {Gondek-Rosi{\'n}ska}, {Gourgoulhon}, {Haensel}, {Taniguchi}, and
  {Zdunik}}}]{Bejger05}
\bibinfo{author}{\bibfnamefont{M.}~\bibnamefont{{Bejger}}},
  \bibinfo{author}{\bibfnamefont{D.}~\bibnamefont{{Gondek-Rosi{\'n}ska}}},
  \bibinfo{author}{\bibfnamefont{E.}~\bibnamefont{{Gourgoulhon}}},
  \bibinfo{author}{\bibfnamefont{P.}~\bibnamefont{{Haensel}}},
  \bibinfo{author}{\bibfnamefont{K.}~\bibnamefont{{Taniguchi}}},
  \bibnamefont{and} \bibinfo{author}{\bibfnamefont{J.~L.}
  \bibnamefont{{Zdunik}}}, \bibinfo{journal}{Astron. Astroph.}
  \textbf{\bibinfo{volume}{431}}, \bibinfo{pages}{297} (\bibinfo{year}{2005}).

\bibitem[{\citenamefont{Vallisneri}(2000)}]{Vallisneri00}
\bibinfo{author}{\bibfnamefont{M.}~\bibnamefont{Vallisneri}},
  \bibinfo{journal}{Phys. Rev. Lett.} \textbf{\bibinfo{volume}{84}},
  \bibinfo{pages}{3519} (\bibinfo{year}{2000}).

\bibitem[{\citenamefont{Narayan et~al.}(1992)\citenamefont{Narayan, Paczynski,
  and Piran}}]{Narayan92}
\bibinfo{author}{\bibfnamefont{R.}~\bibnamefont{Narayan}},
  \bibinfo{author}{\bibfnamefont{B.}~\bibnamefont{Paczynski}},
  \bibnamefont{and} \bibinfo{author}{\bibfnamefont{T.}~\bibnamefont{Piran}},
  \bibinfo{journal}{Astrophys. J.} \textbf{\bibinfo{volume}{395}},
  \bibinfo{pages}{L83} (\bibinfo{year}{1992}).

\bibitem[{\citenamefont{Fox et~al.}(2005)\citenamefont{Fox, Frail, Price,
  Kulkarni, Berger, Piran, Soderberg, Cenko, Cameron, Gal-Yam
  et~al.}}]{Fox2005}
\bibinfo{author}{\bibfnamefont{D.~B.} \bibnamefont{Fox}},
  \bibinfo{author}{\bibfnamefont{D.~A.} \bibnamefont{Frail}},
  \bibinfo{author}{\bibfnamefont{P.~A.} \bibnamefont{Price}},
  \bibinfo{author}{\bibfnamefont{S.~R.} \bibnamefont{Kulkarni}},
  \bibinfo{author}{\bibfnamefont{E.}~\bibnamefont{Berger}},
  \bibinfo{author}{\bibfnamefont{T.}~\bibnamefont{Piran}},
  \bibinfo{author}{\bibfnamefont{A.~M.} \bibnamefont{Soderberg}},
  \bibinfo{author}{\bibfnamefont{S.~B.} \bibnamefont{Cenko}},
  \bibinfo{author}{\bibfnamefont{P.~B.} \bibnamefont{Cameron}},
  \bibinfo{author}{\bibfnamefont{A.}~\bibnamefont{Gal-Yam}},
  \bibnamefont{et~al.}, \bibinfo{journal}{Nature}
  \textbf{\bibinfo{volume}{437}}, \bibinfo{pages}{845} (\bibinfo{year}{2005}).

\bibitem[{\citenamefont{{Kouveliotou} et~al.}(1993)\citenamefont{{Kouveliotou},
  {Meegan}, {Fishman}, {Bhat}, {Briggs}, {Koshut}, {Paciesas}, and
  {Pendleton}}}]{Kouveliotou1993}
\bibinfo{author}{\bibfnamefont{C.}~\bibnamefont{{Kouveliotou}}},
  \bibinfo{author}{\bibfnamefont{C.~A.} \bibnamefont{{Meegan}}},
  \bibinfo{author}{\bibfnamefont{G.~J.} \bibnamefont{{Fishman}}},
  \bibinfo{author}{\bibfnamefont{N.~P.} \bibnamefont{{Bhat}}},
  \bibinfo{author}{\bibfnamefont{M.~S.} \bibnamefont{{Briggs}}},
  \bibinfo{author}{\bibfnamefont{T.~M.} \bibnamefont{{Koshut}}},
  \bibinfo{author}{\bibfnamefont{W.~S.} \bibnamefont{{Paciesas}}},
  \bibnamefont{and} \bibinfo{author}{\bibfnamefont{G.~N.}
  \bibnamefont{{Pendleton}}}, \bibinfo{journal}{Astrophys. J.}
  \textbf{\bibinfo{volume}{413}}, \bibinfo{pages}{L101} (\bibinfo{year}{1993}).

\bibitem[{\citenamefont{{Portegies Zwart}}(1998)}]{Zwart1998a}
\bibinfo{author}{\bibfnamefont{S.~F.} \bibnamefont{{Portegies Zwart}}},
  \bibinfo{journal}{Astrophys. J. Lett.} \textbf{\bibinfo{volume}{503}},
  \bibinfo{pages}{L53} (\bibinfo{year}{1998}).

\bibitem[{\citenamefont{Lee and Kluzniak}(1999{\natexlab{a}})}]{Lee99a}
\bibinfo{author}{\bibfnamefont{W.~H.} \bibnamefont{Lee}} \bibnamefont{and}
  \bibinfo{author}{\bibfnamefont{W.}~\bibnamefont{Kluzniak}},
  \bibinfo{journal}{MNRAS} \textbf{\bibinfo{volume}{308}}, \bibinfo{pages}{780}
  (\bibinfo{year}{1999}{\natexlab{a}}).

\bibitem[{\citenamefont{Lee and Kluzniak}(1999{\natexlab{b}})}]{Lee99b}
\bibinfo{author}{\bibfnamefont{W.~H.} \bibnamefont{Lee}} \bibnamefont{and}
  \bibinfo{author}{\bibfnamefont{W.}~\bibnamefont{Kluzniak}},
  \bibinfo{journal}{Astrophys. J.} \textbf{\bibinfo{volume}{526}},
  \bibinfo{pages}{178} (\bibinfo{year}{1999}{\natexlab{b}}).

\bibitem[{\citenamefont{Miller}(2005)}]{Miller05}
\bibinfo{author}{\bibfnamefont{M.~C.} \bibnamefont{Miller}},
  \bibinfo{journal}{Astrophys. J.} \textbf{\bibinfo{volume}{626}},
  \bibinfo{pages}{L41} (\bibinfo{year}{2005}).

\bibitem[{\citenamefont{Campanelli et~al.}(2006)\citenamefont{Campanelli,
  Lousto, and Zlochower}}]{Campanelli:2006gf}
\bibinfo{author}{\bibfnamefont{M.}~\bibnamefont{Campanelli}},
  \bibinfo{author}{\bibfnamefont{C.~O.} \bibnamefont{Lousto}},
  \bibnamefont{and}
  \bibinfo{author}{\bibfnamefont{Y.}~\bibnamefont{Zlochower}},
  \bibinfo{journal}{Phys. Rev. D} \textbf{\bibinfo{volume}{73}},
  \bibinfo{pages}{061501(R)} (\bibinfo{year}{2006}).

\bibitem[{\citenamefont{Diener et~al.}(2006)\citenamefont{Diener, Herrmann,
  Pollney, Schnetter, Seidel, Takahashi, Thornburg, and
  Ventrella}}]{Diener-etal-2006a}
\bibinfo{author}{\bibfnamefont{P.}~\bibnamefont{Diener}},
  \bibinfo{author}{\bibfnamefont{F.}~\bibnamefont{Herrmann}},
  \bibinfo{author}{\bibfnamefont{D.}~\bibnamefont{Pollney}},
  \bibinfo{author}{\bibfnamefont{E.}~\bibnamefont{Schnetter}},
  \bibinfo{author}{\bibfnamefont{E.}~\bibnamefont{Seidel}},
  \bibinfo{author}{\bibfnamefont{R.}~\bibnamefont{Takahashi}},
  \bibinfo{author}{\bibfnamefont{J.}~\bibnamefont{Thornburg}},
  \bibnamefont{and}
  \bibinfo{author}{\bibfnamefont{J.}~\bibnamefont{Ventrella}},
  \bibinfo{journal}{Phys. Rev. Lett.} \textbf{\bibinfo{volume}{96}},
  \bibinfo{pages}{121101} (\bibinfo{year}{2006}).

\bibitem[{\citenamefont{Baker et~al.}(2006)\citenamefont{Baker, Centrella,
  Choi, Koppitz, and van Meter}}]{Baker:2006yw}
\bibinfo{author}{\bibfnamefont{J.~G.} \bibnamefont{Baker}},
  \bibinfo{author}{\bibfnamefont{J.}~\bibnamefont{Centrella}},
  \bibinfo{author}{\bibfnamefont{D.-I.} \bibnamefont{Choi}},
  \bibinfo{author}{\bibfnamefont{M.}~\bibnamefont{Koppitz}}, \bibnamefont{and}
  \bibinfo{author}{\bibfnamefont{J.}~\bibnamefont{van Meter}},
  \bibinfo{journal}{Phys. Rev. D} \textbf{\bibinfo{volume}{73}},
  \bibinfo{pages}{104002} (\bibinfo{year}{2006}).

\bibitem[{\citenamefont{Shibata et~al.}(2003)\citenamefont{Shibata, Taniguchi,
  and Ury{\={u}}}}]{Shibata:2003ga}
\bibinfo{author}{\bibfnamefont{M.}~\bibnamefont{Shibata}},
  \bibinfo{author}{\bibfnamefont{K.}~\bibnamefont{Taniguchi}},
  \bibnamefont{and}
  \bibinfo{author}{\bibfnamefont{K.}~\bibnamefont{Ury{\={u}}}},
  \bibinfo{journal}{Phys. Rev. D} \textbf{\bibinfo{volume}{68}},
  \bibinfo{pages}{084020} (\bibinfo{year}{2003}).

\bibitem[{\citenamefont{{Shibata} and {Taniguchi}}(2006)}]{Shibata06a}
\bibinfo{author}{\bibfnamefont{M.}~\bibnamefont{{Shibata}}} \bibnamefont{and}
  \bibinfo{author}{\bibfnamefont{K.}~\bibnamefont{{Taniguchi}}},
  \bibinfo{journal}{Phys. Rev. D} \textbf{\bibinfo{volume}{73}},
  \bibinfo{pages}{{064027}} (\bibinfo{year}{2006}).

\bibitem[{\citenamefont{Janka et~al.}(1999)\citenamefont{Janka, Eberl, Ruffert,
  and Fryer}}]{Janka99a}
\bibinfo{author}{\bibfnamefont{H.-T.} \bibnamefont{Janka}},
  \bibinfo{author}{\bibfnamefont{T.}~\bibnamefont{Eberl}},
  \bibinfo{author}{\bibfnamefont{M.}~\bibnamefont{Ruffert}}, \bibnamefont{and}
  \bibinfo{author}{\bibfnamefont{C.~L.} \bibnamefont{Fryer}},
  \bibinfo{journal}{Astrophys. J.} \textbf{\bibinfo{volume}{527}},
  \bibinfo{pages}{L39} (\bibinfo{year}{1999}).

\bibitem[{\citenamefont{Rosswog}(2005)}]{Rosswog05}
\bibinfo{author}{\bibfnamefont{S.}~\bibnamefont{Rosswog}},
  \bibinfo{journal}{Astrophys. J.} \textbf{\bibinfo{volume}{634}},
  \bibinfo{pages}{1202} (\bibinfo{year}{2005}).

\bibitem[{\citenamefont{Baumgarte et~al.}(2004)\citenamefont{Baumgarte, Skoge,
  and Shapiro}}]{Baumgarte:2004}
\bibinfo{author}{\bibfnamefont{T.~W.} \bibnamefont{Baumgarte}},
  \bibinfo{author}{\bibfnamefont{M.~L.} \bibnamefont{Skoge}}, \bibnamefont{and}
  \bibinfo{author}{\bibfnamefont{S.~L.} \bibnamefont{Shapiro}},
  \bibinfo{journal}{Phys. Rev. D} \textbf{\bibinfo{volume}{70}},
  \bibinfo{pages}{064040} (\bibinfo{year}{2004}).

\bibitem[{\citenamefont{Alcubierre et~al.}(2005)\citenamefont{Alcubierre,
  Br{\"u}gmann, Diener, Guzm{\'a}n, Hawke, Hawley, Herrmann, Koppitz, Pollney,
  Seidel et~al.}}]{Alcubierre2003:pre-ISCO-coalescence-times}
\bibinfo{author}{\bibfnamefont{M.}~\bibnamefont{Alcubierre}},
  \bibinfo{author}{\bibfnamefont{B.}~\bibnamefont{Br{\"u}gmann}},
  \bibinfo{author}{\bibfnamefont{P.}~\bibnamefont{Diener}},
  \bibinfo{author}{\bibfnamefont{F.~S.} \bibnamefont{Guzm{\'a}n}},
  \bibinfo{author}{\bibfnamefont{I.}~\bibnamefont{Hawke}},
  \bibinfo{author}{\bibfnamefont{S.}~\bibnamefont{Hawley}},
  \bibinfo{author}{\bibfnamefont{F.}~\bibnamefont{Herrmann}},
  \bibinfo{author}{\bibfnamefont{M.}~\bibnamefont{Koppitz}},
  \bibinfo{author}{\bibfnamefont{D.}~\bibnamefont{Pollney}},
  \bibinfo{author}{\bibfnamefont{E.}~\bibnamefont{Seidel}},
  \bibnamefont{et~al.}, \bibinfo{journal}{Phys. Rev. D}
  \textbf{\bibinfo{volume}{72}}, \bibinfo{pages}{044004}
  (\bibinfo{year}{2005}).

\bibitem[{\citenamefont{Miller et~al.}(2004)\citenamefont{Miller, Gressman, and
  Suen}}]{Miller04}
\bibinfo{author}{\bibfnamefont{M.}~\bibnamefont{Miller}},
  \bibinfo{author}{\bibfnamefont{P.}~\bibnamefont{Gressman}}, \bibnamefont{and}
  \bibinfo{author}{\bibfnamefont{W.-M.} \bibnamefont{Suen}},
  \bibinfo{journal}{Phys. Rev. D} \textbf{\bibinfo{volume}{69}},
  \bibinfo{pages}{064026} (\bibinfo{year}{2004}).

\bibitem[{\citenamefont{Bishop et~al.}(2005)\citenamefont{Bishop, G\'{o}mez,
  Lehner, Maharaj, and Winicour}}]{Bishop05}
\bibinfo{author}{\bibfnamefont{N.~T.} \bibnamefont{Bishop}},
  \bibinfo{author}{\bibfnamefont{R.}~\bibnamefont{G\'{o}mez}},
  \bibinfo{author}{\bibfnamefont{L.}~\bibnamefont{Lehner}},
  \bibinfo{author}{\bibfnamefont{M.}~\bibnamefont{Maharaj}}, \bibnamefont{and}
  \bibinfo{author}{\bibfnamefont{J.}~\bibnamefont{Winicour}},
  \bibinfo{journal}{Phys. Rev. D} \textbf{\bibinfo{volume}{72}},
  \bibinfo{pages}{024002} (\bibinfo{year}{2005}).

\bibitem[{\citenamefont{Faber et~al.}(2005)\citenamefont{Faber, Baumgarte,
  Shapiro, Taniguchi, and Rasio}}]{Faber05}
\bibinfo{author}{\bibfnamefont{J.~A.} \bibnamefont{Faber}},
  \bibinfo{author}{\bibfnamefont{T.~W.} \bibnamefont{Baumgarte}},
  \bibinfo{author}{\bibfnamefont{S.~L.} \bibnamefont{Shapiro}},
  \bibinfo{author}{\bibfnamefont{K.}~\bibnamefont{Taniguchi}},
  \bibnamefont{and} \bibinfo{author}{\bibfnamefont{F.~A.} \bibnamefont{Rasio}},
  \bibinfo{journal}{Phys. Rev. D} \textbf{\bibinfo{volume}{73}},
  \bibinfo{pages}{024012} (\bibinfo{year}{2005}).

\bibitem[{\citenamefont{Taniguchi et~al.}(2005)\citenamefont{Taniguchi,
  Baumgarte, Faber, and Shapiro}}]{Taniguchi05}
\bibinfo{author}{\bibfnamefont{K.}~\bibnamefont{Taniguchi}},
  \bibinfo{author}{\bibfnamefont{T.~W.} \bibnamefont{Baumgarte}},
  \bibinfo{author}{\bibfnamefont{J.~A.} \bibnamefont{Faber}}, \bibnamefont{and}
  \bibinfo{author}{\bibfnamefont{S.~L.} \bibnamefont{Shapiro}},
  \bibinfo{journal}{Phys. Rev. D} \textbf{\bibinfo{volume}{72}},
  \bibinfo{pages}{044008} (\bibinfo{year}{2005}).

\bibitem[{\citenamefont{Sopuerta et~al.}(2006)\citenamefont{Sopuerta, Sperhake,
  and Laguna}}]{Sopuerta:2006bw}
\bibinfo{author}{\bibfnamefont{C.~F.} \bibnamefont{Sopuerta}},
  \bibinfo{author}{\bibfnamefont{U.}~\bibnamefont{Sperhake}}, \bibnamefont{and}
  \bibinfo{author}{\bibfnamefont{P.}~\bibnamefont{Laguna}},
  \bibinfo{journal}{to appear in a special issue of Classical and Quantum
  Gravity}  (\bibinfo{year}{2006}).

\bibitem[{\citenamefont{Baiotti et~al.}(2003)\citenamefont{Baiotti, Hawke,
  Montero, and Rezzolla}}]{Baiotti03a}
\bibinfo{author}{\bibfnamefont{L.}~\bibnamefont{Baiotti}},
  \bibinfo{author}{\bibfnamefont{I.}~\bibnamefont{Hawke}},
  \bibinfo{author}{\bibfnamefont{P.}~\bibnamefont{Montero}}, \bibnamefont{and}
  \bibinfo{author}{\bibfnamefont{L.}~\bibnamefont{Rezzolla}}, in
  \emph{\bibinfo{booktitle}{Computational Astrophysics in Italy: Methods and
  Tools}}, edited by
  \bibinfo{editor}{\bibfnamefont{R.}~\bibnamefont{Capuzzo-Dolcetta}}
  (\bibinfo{publisher}{Mem. Soc. Astron. It. Suppl.},
  \bibinfo{address}{Trieste}, \bibinfo{year}{2003}), vol.~\bibinfo{volume}{1},
  p. \bibinfo{pages}{327}.

\bibitem[{\citenamefont{Baiotti et~al.}(2005)\citenamefont{Baiotti, Hawke,
  Montero, L{\"o}ffler, Rezzolla, Stergioulas, Font, and Seidel}}]{Baiotti04}
\bibinfo{author}{\bibfnamefont{L.}~\bibnamefont{Baiotti}},
  \bibinfo{author}{\bibfnamefont{I.}~\bibnamefont{Hawke}},
  \bibinfo{author}{\bibfnamefont{P.~J.} \bibnamefont{Montero}},
  \bibinfo{author}{\bibfnamefont{F.}~\bibnamefont{L{\"o}ffler}},
  \bibinfo{author}{\bibfnamefont{L.}~\bibnamefont{Rezzolla}},
  \bibinfo{author}{\bibfnamefont{N.}~\bibnamefont{Stergioulas}},
  \bibinfo{author}{\bibfnamefont{J.~A.} \bibnamefont{Font}}, \bibnamefont{and}
  \bibinfo{author}{\bibfnamefont{E.}~\bibnamefont{Seidel}},
  \bibinfo{journal}{Phys. Rev. D} \textbf{\bibinfo{volume}{71}},
  \bibinfo{pages}{024035} (\bibinfo{year}{2005}).

\bibitem[{\citenamefont{Alcubierre et~al.}(2000)\citenamefont{Alcubierre,
  Br\"{u}gmann, Dramlitsch, Font, Papadopoulos, Seidel, Stergioulas, and
  Takahashi}}]{Alcubierre99d}
\bibinfo{author}{\bibfnamefont{M.}~\bibnamefont{Alcubierre}},
  \bibinfo{author}{\bibfnamefont{B.}~\bibnamefont{Br\"{u}gmann}},
  \bibinfo{author}{\bibfnamefont{T.}~\bibnamefont{Dramlitsch}},
  \bibinfo{author}{\bibfnamefont{J.~A.} \bibnamefont{Font}},
  \bibinfo{author}{\bibfnamefont{P.}~\bibnamefont{Papadopoulos}},
  \bibinfo{author}{\bibfnamefont{E.}~\bibnamefont{Seidel}},
  \bibinfo{author}{\bibfnamefont{N.}~\bibnamefont{Stergioulas}},
  \bibnamefont{and}
  \bibinfo{author}{\bibfnamefont{R.}~\bibnamefont{Takahashi}},
  \bibinfo{journal}{Phys. Rev. D} \textbf{\bibinfo{volume}{62}},
  \bibinfo{pages}{044034} (\bibinfo{year}{2000}).

\bibitem[{\citenamefont{Arnowitt et~al.}(1962)\citenamefont{Arnowitt, Deser,
  and Misner}}]{Arnowitt62}
\bibinfo{author}{\bibfnamefont{R.}~\bibnamefont{Arnowitt}},
  \bibinfo{author}{\bibfnamefont{S.}~\bibnamefont{Deser}}, \bibnamefont{and}
  \bibinfo{author}{\bibfnamefont{C.~W.} \bibnamefont{Misner}}, in
  \emph{\bibinfo{booktitle}{Gravitation: An introduction to current research}},
  edited by \bibinfo{editor}{\bibfnamefont{L.}~\bibnamefont{Witten}}
  (\bibinfo{publisher}{John Wiley}, \bibinfo{address}{New York},
  \bibinfo{year}{1962}), pp. \bibinfo{pages}{227--265}.

\bibitem[{\citenamefont{Nakamura et~al.}(1987)\citenamefont{Nakamura, Oohara,
  and Kojima}}]{Nakamura87}
\bibinfo{author}{\bibfnamefont{T.}~\bibnamefont{Nakamura}},
  \bibinfo{author}{\bibfnamefont{K.}~\bibnamefont{Oohara}}, \bibnamefont{and}
  \bibinfo{author}{\bibfnamefont{Y.}~\bibnamefont{Kojima}},
  \bibinfo{journal}{Prog. Theor. Phys. Suppl.} \textbf{\bibinfo{volume}{90}},
  \bibinfo{pages}{1} (\bibinfo{year}{1987}).

\bibitem[{\citenamefont{Shibata and Nakamura}(1995)}]{Shibata95}
\bibinfo{author}{\bibfnamefont{M.}~\bibnamefont{Shibata}} \bibnamefont{and}
  \bibinfo{author}{\bibfnamefont{T.}~\bibnamefont{Nakamura}},
  \bibinfo{journal}{Phys. Rev. D} \textbf{\bibinfo{volume}{52}},
  \bibinfo{pages}{5428} (\bibinfo{year}{1995}).

\bibitem[{\citenamefont{Alcubierre et~al.}(2003)\citenamefont{Alcubierre,
  Br\"ugmann, Diener, Koppitz, Pollney, Seidel, and Takahashi}}]{Alcubierre02a}
\bibinfo{author}{\bibfnamefont{M.}~\bibnamefont{Alcubierre}},
  \bibinfo{author}{\bibfnamefont{B.}~\bibnamefont{Br\"ugmann}},
  \bibinfo{author}{\bibfnamefont{P.}~\bibnamefont{Diener}},
  \bibinfo{author}{\bibfnamefont{M.}~\bibnamefont{Koppitz}},
  \bibinfo{author}{\bibfnamefont{D.}~\bibnamefont{Pollney}},
  \bibinfo{author}{\bibfnamefont{E.}~\bibnamefont{Seidel}}, \bibnamefont{and}
  \bibinfo{author}{\bibfnamefont{R.}~\bibnamefont{Takahashi}},
  \bibinfo{journal}{Phys. Rev. D} \textbf{\bibinfo{volume}{67}},
  \bibinfo{pages}{084023} (\bibinfo{year}{2003}).

\bibitem[{\citenamefont{Bona et~al.}(2004)\citenamefont{Bona, Ledvinka,
  Palenzuela, and Zacek}}]{Bona:2003qn}
\bibinfo{author}{\bibfnamefont{C.}~\bibnamefont{Bona}},
  \bibinfo{author}{\bibfnamefont{T.}~\bibnamefont{Ledvinka}},
  \bibinfo{author}{\bibfnamefont{C.}~\bibnamefont{Palenzuela}},
  \bibnamefont{and} \bibinfo{author}{\bibfnamefont{M.}~\bibnamefont{Zacek}},
  \bibinfo{journal}{Phys. Rev. D} \textbf{\bibinfo{volume}{69}},
  \bibinfo{pages}{064036} (\bibinfo{year}{2004}).

\bibitem[{\citenamefont{Nagy et~al.}(2004)\citenamefont{Nagy, Ortiz, and
  Reula}}]{Nagy:2004td}
\bibinfo{author}{\bibfnamefont{G.}~\bibnamefont{Nagy}},
  \bibinfo{author}{\bibfnamefont{O.~E.} \bibnamefont{Ortiz}}, \bibnamefont{and}
  \bibinfo{author}{\bibfnamefont{O.~A.} \bibnamefont{Reula}},
  \bibinfo{journal}{Phys. Rev. D} \textbf{\bibinfo{volume}{70}},
  \bibinfo{pages}{044012} (\bibinfo{year}{2004}).

\bibitem[{\citenamefont{Alcubierre et~al.}(2001)\citenamefont{Alcubierre,
  Br\"ugmann, Pollney, Seidel, and Takahashi}}]{Alcubierre01a}
\bibinfo{author}{\bibfnamefont{M.}~\bibnamefont{Alcubierre}},
  \bibinfo{author}{\bibfnamefont{B.}~\bibnamefont{Br\"ugmann}},
  \bibinfo{author}{\bibfnamefont{D.}~\bibnamefont{Pollney}},
  \bibinfo{author}{\bibfnamefont{E.}~\bibnamefont{Seidel}}, \bibnamefont{and}
  \bibinfo{author}{\bibfnamefont{R.}~\bibnamefont{Takahashi}},
  \bibinfo{journal}{Phys. Rev. D} \textbf{\bibinfo{volume}{64}},
  \bibinfo{pages}{061501(R)} (\bibinfo{year}{2001}).

\bibitem[{\citenamefont{Bona et~al.}(1995)\citenamefont{Bona, Mass{\'o},
  Seidel, and Stela}}]{Bona94b}
\bibinfo{author}{\bibfnamefont{C.}~\bibnamefont{Bona}},
  \bibinfo{author}{\bibfnamefont{J.}~\bibnamefont{Mass{\'o}}},
  \bibinfo{author}{\bibfnamefont{E.}~\bibnamefont{Seidel}}, \bibnamefont{and}
  \bibinfo{author}{\bibfnamefont{J.}~\bibnamefont{Stela}},
  \bibinfo{journal}{Phys. Rev. Lett.} \textbf{\bibinfo{volume}{75}},
  \bibinfo{pages}{600} (\bibinfo{year}{1995}).

\bibitem[{\citenamefont{Alcubierre and Mass{\'o}}(1998)}]{Alcubierre97b}
\bibinfo{author}{\bibfnamefont{M.}~\bibnamefont{Alcubierre}} \bibnamefont{and}
  \bibinfo{author}{\bibfnamefont{J.}~\bibnamefont{Mass{\'o}}},
  \bibinfo{journal}{Phys. Rev. D} \textbf{\bibinfo{volume}{57}},
  \bibinfo{pages}{4511} (\bibinfo{year}{1998}).

\bibitem[{\citenamefont{Smarr and York}(1978)}]{Smarr78b}
\bibinfo{author}{\bibfnamefont{L.}~\bibnamefont{Smarr}} \bibnamefont{and}
  \bibinfo{author}{\bibfnamefont{J.~W.} \bibnamefont{York}},
  \bibinfo{journal}{Phys. Rev. D} \textbf{\bibinfo{volume}{17}},
  \bibinfo{pages}{2529} (\bibinfo{year}{1978}).

\bibitem[{\citenamefont{Mart\'{\i} et~al.}(1991)\citenamefont{Mart\'{\i},
  Ib{\'a}{\~n}ez, and Miralles}}]{Marti91}
\bibinfo{author}{\bibfnamefont{J.~M.} \bibnamefont{Mart\'{\i}}},
  \bibinfo{author}{\bibfnamefont{J.~M.} \bibnamefont{Ib{\'a}{\~n}ez}},
  \bibnamefont{and} \bibinfo{author}{\bibfnamefont{J.~M.}
  \bibnamefont{Miralles}}, \bibinfo{journal}{Phys. Rev. D}
  \textbf{\bibinfo{volume}{43}}, \bibinfo{pages}{3794} (\bibinfo{year}{1991}).

\bibitem[{\citenamefont{Banyuls et~al.}(1997)\citenamefont{Banyuls, Font,
  Ib{\'a}{\~n}ez, Mart\'{\i}, and Miralles}}]{Banyuls97}
\bibinfo{author}{\bibfnamefont{F.}~\bibnamefont{Banyuls}},
  \bibinfo{author}{\bibfnamefont{J.~A.} \bibnamefont{Font}},
  \bibinfo{author}{\bibfnamefont{J.~M.} \bibnamefont{Ib{\'a}{\~n}ez}},
  \bibinfo{author}{\bibfnamefont{J.~M.} \bibnamefont{Mart\'{\i}}},
  \bibnamefont{and} \bibinfo{author}{\bibfnamefont{J.~A.}
  \bibnamefont{Miralles}}, \bibinfo{journal}{Astrophys. J.}
  \textbf{\bibinfo{volume}{476}}, \bibinfo{pages}{221} (\bibinfo{year}{1997}).

\bibitem[{\citenamefont{Ib{\'a}{\~n}ez
  et~al.}(2001)\citenamefont{Ib{\'a}{\~n}ez, Aloy, Font, Mart\'{\i}, Miralles,
  and Pons}}]{Ibanez01}
\bibinfo{author}{\bibfnamefont{J.}~\bibnamefont{Ib{\'a}{\~n}ez}},
  \bibinfo{author}{\bibfnamefont{M.}~\bibnamefont{Aloy}},
  \bibinfo{author}{\bibfnamefont{J.}~\bibnamefont{Font}},
  \bibinfo{author}{\bibfnamefont{J.}~\bibnamefont{Mart\'{\i}}},
  \bibinfo{author}{\bibfnamefont{J.}~\bibnamefont{Miralles}}, \bibnamefont{and}
  \bibinfo{author}{\bibfnamefont{J.}~\bibnamefont{Pons}}, in
  \emph{\bibinfo{booktitle}{Godunov methods: theory and applications}}, edited
  by \bibinfo{editor}{\bibfnamefont{E.}~\bibnamefont{Toro}}
  (\bibinfo{publisher}{Kluwer Academic/Plenum Publishers},
  \bibinfo{address}{New York}, \bibinfo{year}{2001}).

\bibitem[{\citenamefont{Font}(2003)}]{Font03}
\bibinfo{author}{\bibfnamefont{J.~A.} \bibnamefont{Font}},
  \bibinfo{journal}{Liv. Rev. Relativ.} \textbf{\bibinfo{volume}{6}},
  \bibinfo{pages}{4} (\bibinfo{year}{2003}),
  \bibinfo{note}{http://www.livingreviews.org/lrr-2003-4}.

\bibitem[{\citenamefont{Ansorg et~al.}(2004)\citenamefont{Ansorg, Br\"ugmann,
  and Tichy}}]{Ansorg:2004ds}
\bibinfo{author}{\bibfnamefont{M.}~\bibnamefont{Ansorg}},
  \bibinfo{author}{\bibfnamefont{B.}~\bibnamefont{Br\"ugmann}},
  \bibnamefont{and} \bibinfo{author}{\bibfnamefont{W.}~\bibnamefont{Tichy}},
  \bibinfo{journal}{Phys. Rev. D} \textbf{\bibinfo{volume}{70}},
  \bibinfo{pages}{064011} (\bibinfo{year}{2004}).

\bibitem[{\citenamefont{Brandt et~al.}(2000)\citenamefont{Brandt, Correll,
  G\'{o}mez, Huq, Laguna, Lehner, Marronetti, Matzner, Neilsen, Pullin
  et~al.}}]{Brandt00}
\bibinfo{author}{\bibfnamefont{S.}~\bibnamefont{Brandt}},
  \bibinfo{author}{\bibfnamefont{R.}~\bibnamefont{Correll}},
  \bibinfo{author}{\bibfnamefont{R.}~\bibnamefont{G\'{o}mez}},
  \bibinfo{author}{\bibfnamefont{M.~F.} \bibnamefont{Huq}},
  \bibinfo{author}{\bibfnamefont{P.}~\bibnamefont{Laguna}},
  \bibinfo{author}{\bibfnamefont{L.}~\bibnamefont{Lehner}},
  \bibinfo{author}{\bibfnamefont{P.}~\bibnamefont{Marronetti}},
  \bibinfo{author}{\bibfnamefont{R.~A.} \bibnamefont{Matzner}},
  \bibinfo{author}{\bibfnamefont{D.}~\bibnamefont{Neilsen}},
  \bibinfo{author}{\bibfnamefont{J.}~\bibnamefont{Pullin}},
  \bibnamefont{et~al.}, \bibinfo{journal}{Phys. Rev. Lett.}
  \textbf{\bibinfo{volume}{85}}, \bibinfo{pages}{5496} (\bibinfo{year}{2000}).

\bibitem[{\citenamefont{York}(1979)}]{York79}
\bibinfo{author}{\bibfnamefont{J.~W.} \bibnamefont{York}}, in
  \emph{\bibinfo{booktitle}{Sources of gravitational radiation}}, edited by
  \bibinfo{editor}{\bibfnamefont{L.~L.} \bibnamefont{Smarr}}
  (\bibinfo{publisher}{Cambridge University Press},
  \bibinfo{address}{Cambridge, UK}, \bibinfo{year}{1979}), pp.
  \bibinfo{pages}{83--126}, ISBN \bibinfo{isbn}{0-521-22778-X}.

\bibitem[{\citenamefont{Berger}(1982)}]{Berger-1982}
\bibinfo{author}{\bibfnamefont{M.~J.} \bibnamefont{Berger}}, Ph.D. thesis,
  \bibinfo{school}{Stanford University} (\bibinfo{year}{1982}).

\bibitem[{\citenamefont{Schnetter et~al.}(2004)\citenamefont{Schnetter, Hawley,
  and Hawke}}]{Schnetter-etal-03b}
\bibinfo{author}{\bibfnamefont{E.}~\bibnamefont{Schnetter}},
  \bibinfo{author}{\bibfnamefont{S.~H.} \bibnamefont{Hawley}},
  \bibnamefont{and} \bibinfo{author}{\bibfnamefont{I.}~\bibnamefont{Hawke}},
  \bibinfo{journal}{Class. Quantum Grav.} \textbf{\bibinfo{volume}{21}},
  \bibinfo{pages}{1465} (\bibinfo{year}{2004}).

\bibitem[{\citenamefont{Hawke et~al.}(2005)\citenamefont{Hawke, L{\"o}ffler,
  and Nerozzi}}]{Hawke04}
\bibinfo{author}{\bibfnamefont{I.}~\bibnamefont{Hawke}},
  \bibinfo{author}{\bibfnamefont{F.}~\bibnamefont{L{\"o}ffler}},
  \bibnamefont{and} \bibinfo{author}{\bibfnamefont{A.}~\bibnamefont{Nerozzi}},
  \bibinfo{journal}{Phys. Rev. D} \textbf{\bibinfo{volume}{71}},
  \bibinfo{pages}{104006} (\bibinfo{year}{2005}).

\bibitem[{\citenamefont{Kreiss and Oliger}(1973)}]{Kreiss73}
\bibinfo{author}{\bibfnamefont{H.-O.} \bibnamefont{Kreiss}} \bibnamefont{and}
  \bibinfo{author}{\bibfnamefont{J.}~\bibnamefont{Oliger}},
  \bibinfo{journal}{Global atmospheric research programme publications series}
  \textbf{\bibinfo{volume}{10}} (\bibinfo{year}{1973}).

\bibitem[{\citenamefont{Thornburg}(1996)}]{Thornburg95}
\bibinfo{author}{\bibfnamefont{J.}~\bibnamefont{Thornburg}},
  \bibinfo{journal}{Phys. Rev. D} \textbf{\bibinfo{volume}{54}},
  \bibinfo{pages}{4899} (\bibinfo{year}{1996}).

\bibitem[{\citenamefont{Thornburg}(2004{\natexlab{a}})}]{Thornburg2003:AH-find%
ing}
\bibinfo{author}{\bibfnamefont{J.}~\bibnamefont{Thornburg}},
  \bibinfo{journal}{Class. Quantum Grav.} \textbf{\bibinfo{volume}{21}},
  \bibinfo{pages}{743} (\bibinfo{year}{2004}{\natexlab{a}}).

\bibitem[{\citenamefont{Calabrese and
  Neilsen}(2004)}]{Calabrese2003:excision-and-summation-by-parts}
\bibinfo{author}{\bibfnamefont{G.}~\bibnamefont{Calabrese}} \bibnamefont{and}
  \bibinfo{author}{\bibfnamefont{D.}~\bibnamefont{Neilsen}},
  \bibinfo{journal}{Phys. Rev. D} \textbf{\bibinfo{volume}{69}},
  \bibinfo{pages}{044020} (\bibinfo{year}{2004}).

\bibitem[{\citenamefont{Pretorius}(2005)}]{Pretorius05}
\bibinfo{author}{\bibfnamefont{F.}~\bibnamefont{Pretorius}},
  \bibinfo{journal}{Class.Quant.Grav.} \textbf{\bibinfo{volume}{22}},
  \bibinfo{pages}{425} (\bibinfo{year}{2005}).

\bibitem[{\citenamefont{Sperhake et~al.}(2005)\citenamefont{Sperhake, Kelly,
  Laguna, Smith, and Schnetter}}]{Sperhake2005a}
\bibinfo{author}{\bibfnamefont{U.}~\bibnamefont{Sperhake}},
  \bibinfo{author}{\bibfnamefont{B.}~\bibnamefont{Kelly}},
  \bibinfo{author}{\bibfnamefont{P.}~\bibnamefont{Laguna}},
  \bibinfo{author}{\bibfnamefont{K.~L.} \bibnamefont{Smith}}, \bibnamefont{and}
  \bibinfo{author}{\bibfnamefont{E.}~\bibnamefont{Schnetter}},
  \bibinfo{journal}{Phys. Rev. D} \textbf{\bibinfo{volume}{71}},
  \bibinfo{pages}{124042} (\bibinfo{year}{2005}).

\bibitem[{\citenamefont{Anninos et~al.}(1995)\citenamefont{Anninos, Daues,
  Mass{\'o}, Seidel, and Suen}}]{Anninos94e}
\bibinfo{author}{\bibfnamefont{P.}~\bibnamefont{Anninos}},
  \bibinfo{author}{\bibfnamefont{G.}~\bibnamefont{Daues}},
  \bibinfo{author}{\bibfnamefont{J.}~\bibnamefont{Mass{\'o}}},
  \bibinfo{author}{\bibfnamefont{E.}~\bibnamefont{Seidel}}, \bibnamefont{and}
  \bibinfo{author}{\bibfnamefont{W.-M.} \bibnamefont{Suen}},
  \bibinfo{journal}{Phys. Rev. D} \textbf{\bibinfo{volume}{51}},
  \bibinfo{pages}{5562} (\bibinfo{year}{1995}).

\bibitem[{\citenamefont{Br\"ugmann et~al.}(2004)\citenamefont{Br\"ugmann,
  Tichy, and Jansen}}]{Bruegmann:2003aw}
\bibinfo{author}{\bibfnamefont{B.}~\bibnamefont{Br\"ugmann}},
  \bibinfo{author}{\bibfnamefont{W.}~\bibnamefont{Tichy}}, \bibnamefont{and}
  \bibinfo{author}{\bibfnamefont{N.}~\bibnamefont{Jansen}},
  \bibinfo{journal}{Phys. Rev. Lett.} \textbf{\bibinfo{volume}{92}},
  \bibinfo{pages}{211101} (\bibinfo{year}{2004}).

\bibitem[{\citenamefont{{Lin} et~al.}(2006)\citenamefont{{Lin}, {Cheng}, {Chu},
  and {Suen}}}]{lin_etal_06}
\bibinfo{author}{\bibfnamefont{L.-M.} \bibnamefont{{Lin}}},
  \bibinfo{author}{\bibfnamefont{K.~S.} \bibnamefont{{Cheng}}},
  \bibinfo{author}{\bibfnamefont{M.-C.} \bibnamefont{{Chu}}}, \bibnamefont{and}
  \bibinfo{author}{\bibfnamefont{W.-M.} \bibnamefont{{Suen}}},
  \bibinfo{journal}{\apj} \textbf{\bibinfo{volume}{639}}, \bibinfo{pages}{382}
  (\bibinfo{year}{2006}).

\bibitem[{\citenamefont{Leaver}(1985)}]{Leaver85}
\bibinfo{author}{\bibfnamefont{E.}~\bibnamefont{Leaver}},
  \bibinfo{journal}{Proc. R. Soc. London, Ser. A}
  \textbf{\bibinfo{volume}{402}}, \bibinfo{pages}{285} (\bibinfo{year}{1985}).

\bibitem[{\citenamefont{Kokkotas and Schmidt}(1999)}]{Kokkotas99a}
\bibinfo{author}{\bibfnamefont{K.~D.} \bibnamefont{Kokkotas}} \bibnamefont{and}
  \bibinfo{author}{\bibfnamefont{B.~G.} \bibnamefont{Schmidt}},
  \bibinfo{journal}{Living Rev. Rel.} \textbf{\bibinfo{volume}{2}},
  \bibinfo{pages}{1999} (\bibinfo{year}{1999}),
  \bibinfo{note}{http://www.livingreviews.org/lrr-1999-2}.

\bibitem[{\citenamefont{Abrahams et~al.}(1998)\citenamefont{Abrahams, Rezzolla,
  Rupright, Anderson, Anninos, Baumgarte, Bishop, Brandt, Browne, Camarda
  et~al.}}]{Abrahams97a}
\bibinfo{author}{\bibfnamefont{A.~M.} \bibnamefont{Abrahams}},
  \bibinfo{author}{\bibfnamefont{L.}~\bibnamefont{Rezzolla}},
  \bibinfo{author}{\bibfnamefont{M.~E.} \bibnamefont{Rupright}},
  \bibinfo{author}{\bibfnamefont{A.}~\bibnamefont{Anderson}},
  \bibinfo{author}{\bibfnamefont{P.}~\bibnamefont{Anninos}},
  \bibinfo{author}{\bibfnamefont{T.~W.} \bibnamefont{Baumgarte}},
  \bibinfo{author}{\bibfnamefont{N.~T.} \bibnamefont{Bishop}},
  \bibinfo{author}{\bibfnamefont{S.~R.} \bibnamefont{Brandt}},
  \bibinfo{author}{\bibfnamefont{J.~C.} \bibnamefont{Browne}},
  \bibinfo{author}{\bibfnamefont{K.}~\bibnamefont{Camarda}},
  \bibnamefont{et~al.}, \bibinfo{journal}{Phys. Rev. Lett.}
  \textbf{\bibinfo{volume}{80}}, \bibinfo{pages}{1812} (\bibinfo{year}{1998}).

\bibitem[{\citenamefont{Rezzolla et~al.}(1999)\citenamefont{Rezzolla, Abrahams,
  Matzner, Rupright, and Shapiro}}]{Rezzolla99a}
\bibinfo{author}{\bibfnamefont{L.}~\bibnamefont{Rezzolla}},
  \bibinfo{author}{\bibfnamefont{A.~M.} \bibnamefont{Abrahams}},
  \bibinfo{author}{\bibfnamefont{R.~A.} \bibnamefont{Matzner}},
  \bibinfo{author}{\bibfnamefont{M.~E.} \bibnamefont{Rupright}},
  \bibnamefont{and} \bibinfo{author}{\bibfnamefont{S.~L.}
  \bibnamefont{Shapiro}}, \bibinfo{journal}{Phys. Rev. D}
  \textbf{\bibinfo{volume}{59}}, \bibinfo{pages}{064001}
  (\bibinfo{year}{1999}).

\bibitem[{\citenamefont{Calabrese et~al.}(2003)\citenamefont{Calabrese, Pullin,
  Reula, Sarbach, and Tiglio}}]{Calabrese:2002xy}
\bibinfo{author}{\bibfnamefont{G.}~\bibnamefont{Calabrese}},
  \bibinfo{author}{\bibfnamefont{J.}~\bibnamefont{Pullin}},
  \bibinfo{author}{\bibfnamefont{O.}~\bibnamefont{Reula}},
  \bibinfo{author}{\bibfnamefont{O.}~\bibnamefont{Sarbach}}, \bibnamefont{and}
  \bibinfo{author}{\bibfnamefont{M.}~\bibnamefont{Tiglio}},
  \bibinfo{journal}{Communications in Mathematical Physics}
  \textbf{\bibinfo{volume}{240}}, \bibinfo{pages}{377} (\bibinfo{year}{2003}).

\bibitem[{\citenamefont{Gundlach and Martin-Garcia}(2004)}]{Gundlach:2004jp}
\bibinfo{author}{\bibfnamefont{C.}~\bibnamefont{Gundlach}} \bibnamefont{and}
  \bibinfo{author}{\bibfnamefont{J.~M.} \bibnamefont{Martin-Garcia}},
  \bibinfo{journal}{Phys. Rev. D} \textbf{\bibinfo{volume}{70}},
  \bibinfo{pages}{044032} (\bibinfo{year}{2004}).

\bibitem[{\citenamefont{Szilagyi and Winicour}(2003)}]{Szilagyi02a}
\bibinfo{author}{\bibfnamefont{B.}~\bibnamefont{Szilagyi}} \bibnamefont{and}
  \bibinfo{author}{\bibfnamefont{J.}~\bibnamefont{Winicour}},
  \bibinfo{journal}{Phys. Rev. D} \textbf{\bibinfo{volume}{68}},
  \bibinfo{pages}{041501(R)} (\bibinfo{year}{2003}).

\bibitem[{\citenamefont{Rupright et~al.}(1998)\citenamefont{Rupright, Abrahams,
  and Rezzolla}}]{Rupright98}
\bibinfo{author}{\bibfnamefont{M.~E.} \bibnamefont{Rupright}},
  \bibinfo{author}{\bibfnamefont{A.~M.} \bibnamefont{Abrahams}},
  \bibnamefont{and} \bibinfo{author}{\bibfnamefont{L.}~\bibnamefont{Rezzolla}},
  \bibinfo{journal}{Phys. Rev. D} \textbf{\bibinfo{volume}{58}},
  \bibinfo{pages}{044005} (\bibinfo{year}{1998}).

\bibitem[{\citenamefont{Camarda and Seidel}(1999)}]{Camarda97c}
\bibinfo{author}{\bibfnamefont{K.}~\bibnamefont{Camarda}} \bibnamefont{and}
  \bibinfo{author}{\bibfnamefont{E.}~\bibnamefont{Seidel}},
  \bibinfo{journal}{Phys. Rev. D} \textbf{\bibinfo{volume}{59}},
  \bibinfo{pages}{064019} (\bibinfo{year}{1999}).

\bibitem[{\citenamefont{Allen et~al.}(1998)\citenamefont{Allen, Camarda, and
  Seidel}}]{Allen98a}
\bibinfo{author}{\bibfnamefont{G.}~\bibnamefont{Allen}},
  \bibinfo{author}{\bibfnamefont{K.}~\bibnamefont{Camarda}}, \bibnamefont{and}
  \bibinfo{author}{\bibfnamefont{E.}~\bibnamefont{Seidel}}
  (\bibinfo{year}{1998}).

\bibitem[{\citenamefont{Moncrief}(1974)}]{Moncrief74}
\bibinfo{author}{\bibfnamefont{V.}~\bibnamefont{Moncrief}},
  \bibinfo{journal}{Annals of Physics} \textbf{\bibinfo{volume}{88}},
  \bibinfo{pages}{323} (\bibinfo{year}{1974}).

\bibitem[{\citenamefont{L{\"o}ffler}(2005)}]{Loeffler05}
\bibinfo{author}{\bibfnamefont{F.}~\bibnamefont{L{\"o}ffler}}, Ph.D. thesis,
  \bibinfo{school}{Max {P}lanck Institute for Gravitational Physics ({A}lbert
  {E}instein Institute)} (\bibinfo{year}{2005}).

\bibitem[{\citenamefont{Thornburg}(2004{\natexlab{b}})}]{Thornburg2004:multipa%
tch-BH-excision_nourl}
\bibinfo{author}{\bibfnamefont{J.}~\bibnamefont{Thornburg}},
  \bibinfo{journal}{Class. Quantum Grav.} \textbf{\bibinfo{volume}{21}},
  \bibinfo{pages}{3665} (\bibinfo{year}{2004}{\natexlab{b}}).

\bibitem[{\citenamefont{Baiotti and Rezzolla}(2006)}]{Baiotti06}
\bibinfo{author}{\bibfnamefont{L.}~\bibnamefont{Baiotti}} \bibnamefont{and}
  \bibinfo{author}{\bibfnamefont{L.}~\bibnamefont{Rezzolla}},
  \bibinfo{journal}{submitted to Phys. Rev. L.}  (\bibinfo{year}{2006}).

\end{thebibliography}
\end{document}